\documentstyle[preprint,tighten,eqsecnum,floats,aps]{revtex}

\makeatletter

\def\homega{\hat{\omega}}
\def\div{{\rm div}}
\def\d{\delta}

\def\IH{\triangle}
\def\hIH{\hat{\triangle}}
\def\hDelta{\hat{\Delta}}
\def\tIH{\tilde{\IH}}
\def\bIH{\bar{\IH}}

\def\hdiv{\hat{\div}}
\def\q{\hat{q}}

\def\tIH{\tilde{\IH}}

\def\D{{\cal D}}
\def\S{S}
\def\L{{\cal L}}
\def\Lie{\L}
\def\N{N}
\def\I{\mathcal{I}}

\def\const{{\rm const}}

\def\Re{{\rm Re}}
\def\Im{{\rm Im}}
\def\l{\ell}

\def\M{{\bf M}}

\def\be{\begin{equation}}
\def\ee{\end{equation}}
\def\ba{\begin{eqnarray}}
\def\ea{\end{eqnarray}}

\def\R{{\mbox{\rm$\mbox{I}\!\mbox{R}$}}}

\def\pback#1{{
\mathchoice{\StemPullBack{#1}{\leftarrowfill}}
     {\StemPullBack{#1}{\leftarrowfill}}
             {\IndxPullBack{#1}{\leftarrowfill}}
         {\IndxPullBack{#1}{\leftarrowfill}}}\vphantom{#1}}

\newcommand{\StemPullBack}[2]{
  \vtop{\mathsurround=0pt
  \ialign{##\crcr$\textstyle{#1}\strut$\crcr
    \noalign{\kern-0.4ex\nointerlineskip}{\tiny#2}\crcr}}}

\newcommand{\IndxPullBack}[2]{
  \vtop{\mathsurround=0pt
  \ialign{##\crcr\hfil$\scriptstyle{#1}$\hfil\crcr
    \noalign{\kern+0.4ex\nointerlineskip}{\tiny#2}\crcr}}}

\def\a{{\alpha}}

\def\b{{\beta}}
\def\kl{\kappa_{(\ell)}}
\def\klprime{\kappa_{({\ell}')}}
\def\klp{\klprime}

\def\m{\delta}
\def\mb{\bar{\delta}}
\def\bm{\mb}
\def\n{\Delta}
\def\d{\delta}
\newcommand{\eqhat}{\mathrel{\widehat\mathalpha{=}}}
\def\={\eqhat}
\def\g{\gamma}

\def\tepsilon{\tilde{\epsilon}}
\def\tomega{\tilde{\omega}}
\def\tIH{\tilde{\IH}}
\def\tD{\tilde{\D}}
\def\te{\tilde{\epsilon}}
\def\tq{\tilde{q}}

\def\r{(\tilde{r}^{-1})}
\def\tS{\tilde{S}}

\preprint{\vbox{\baselineskip=12pt \rightline{CGPG-01/11-3}
\rightline{gr-qc/0111067} }}

\begin{document}
%\draft
\title{Geometry of Generic Isolated Horizons}
\author {Abhay Ashtekar${}^{1,4}$\thanks{E-mail:
ashtekar@gravity.phys.psu.edu}, Christopher
Beetle${}^{1,2}$\thanks{E-mail: beetle@physics.utah.edu}, and
Jerzy Lewandowski${}^{3,4}$\thanks{E-mail:lewand@fuw.edu.pl} }
\address{{1.} Center for Gravitational Physics and Geometry\\
Physics Department, Penn State, University Park, PA 16802, U.S.A.}
\address{{2.} Physics Department, University of Utah, Salt Lake City,
Utah 84112}
\address{{3.} Institute of Theoretical Physics,
Warsaw University, ul. Hoza 69, 00-681, Warsaw, Poland}
\address{{4.} Max Planck Institut f\"ur Gravitationsphysik, Am M\"uhlenberg 1,
D 14476 Golm, Germany}
\maketitle

\begin{abstract}

Geometrical structures intrinsic to non-expanding, weakly isolated
and isolated horizons are analyzed and compared with structures
which arise in other contexts within general relativity, e.g., at
null infinity. In particular, we address in detail the issue of
singling out the preferred normals to these horizons required in
various applications. This work provides powerful tools to extract
invariant, physical information from numerical simulations of the
near horizon, strong field geometry. While it complements the
previous analysis of laws governing the mechanics of weakly
isolated horizons, prior knowledge of those results is not
assumed.

\end{abstract}
%\pacs{Pacs: 04.70.-s, 04.70.Bw}

\section{Introduction}
\label{s1}

Isolated horizons approximate event horizons of black holes at
late stages of gravitational collapse and of black hole mergers
when back-scattered radiation falling into the hole can be
neglected \cite{prl}. However, unlike event horizons, they are
defined quasi-locally and, unlike Killing horizons, they do not
require the presence of a Killing vector in their neighborhood.
Therefore isolated horizons can be easily located, e.g., in
numerical simulations \cite{num}. They can be rotating and may be
distorted due to the presence of other black holes, matter discs,
external magnetic fields, etc. Consequently, the isolated horizon
framework can serve as a powerful tool in a variety of physical
situations. (For early work in which similar ideas were explored
from somewhat different perspectives, see \cite{tn,ph,td}.)

The notion of an isolated horizon was first introduced to
generalize the laws governing black hole mechanics to more
realistic situations which allow for gravitational (and other)
radiation in the exterior region of space-time
\cite{abf,afk,abl2}. It has since proved to be useful also in
several other contexts, ranging from numerical relativity to
background independent quantum gravity: i) it plays a key role in
an ongoing program for extracting physics from numerical
simulations of black hole mergers \cite{prl,num,abl3}; ii) it has
led to the introduction \cite{cs,afk,acs} of a physical model of
hairy black holes, systematizing a large body of results obtained
from a mixture of analytical and numerical investigations; and,
iii) it serves as a point of departure for statistical mechanical
entropy calculations in which all non-rotating black holes
(extremal or not) \textit{and} cosmological horizons are
incorporated in a single stroke \cite{ack,abck,abk}.

The purpose of this paper is to analyze in detail the
intrinsically defined, geometrical structures on non-expanding,
weakly isolated and isolated horizons and to study their interplay
with Einstein's equations (possibly with matter sources). The key
geometric structures consist of null normals $\l^a$, the intrinsic
(degenerate) metric $q_{ab}$ and the derivative operator $\D$,
induced by the space-time connection $\nabla$. We will analyze
relations among them which follow directly from the definitions of
these horizons; derive the constraints they must satisfy as a
consequence of the pull-back of the field equations to horizons;
specify the ``free data''; and spell out the information about
\textit{space-time} curvature contained in their intrinsic
geometry. In addition, we will address an issue that plays an
important role in applications of the framework, in particular to
numerical relativity \cite{num}: Can we use the intrinsic
structures to select a preferred class of null normals $\l$ to the
horizon? We will show that, generically, the answer is in the
affirmative. Overall, the results of this paper complement those
on mechanics of isolated horizons \cite{abf,afk,abl2}. Whereas in
that work the emphasis was on the infinite dimensional phase space
of space-times admitting isolated horizons as null boundaries, in
the present case we focus on the geometry of isolated horizons in
individual space-times.

In Section \ref{s2} we consider non-expanding horizons (NEHs)
$\IH$. These are 3-dimensional, null sub-manifolds of space-time
with vanishing expansion on which a mild energy condition holds.
We show they are naturally equipped with a (degenerate)
metric $q$ and a derivative operator $\D$ and analyze the restrictions
the space-time curvature must satisfy on them.  In Section
\ref{s3} we define weakly isolated horizons (WIHs) $(\IH, [\l])$
by imposing a restriction on an equivalence class $[\l]$ of future
directed null normals $\l$ to $\IH$ (where two are equivalent if
they are \textit{constant} multiples of each other). The pull-back
of the field equations to $\IH$ constrain the geometrical pair $(q,
\D)$. We analyze these constraint equations and spell out the
freely specifiable parts of these variables. Finally, we show that
in the case when the surface gravity is non-zero, the WIHs $(\IH,
[\l])$ admit a natural foliation which can, in particular, be used
in numerical relativity to construct invariantly defined
coordinates and tetrads. A WIH  has certain similarities with null
infinity since both can be regarded as null boundaries. We show
that the intrinsic geometric structure of WIHs and its interplay
with field equations is rather similar to that at null infinity in
absence of radiation.

A natural question is whether every NEH  $\IH$ admits a null
normal $\l$ such that $(\IH, [\l])$ is a WIH and if the choice is
unique. The answer to the existence question is in the
affirmative. However, the choice is far from being unique. In
Section \ref{s4}, therefore, we consider additional geometrical
conditions on $\l$ which \textit{can} select $[\l]$ uniquely. It
turns out there is an obvious choice which fulfills this task.
Although this choice seems natural at first, in the case when
$\Delta$ {\it is\/} a Killing horizon, the restriction of the
Killing field to the horizon need not belong to the equivalence
class $[\l]$ so chosen. Furthermore, in this case, $[\l]$ need not
even be preserved by the isometry generated by the Killing field.
We then assume that the pull-back of the space-time Ricci tensor
to $\IH$ satisfies a natural condition and introduce a more
sophisticated restriction on $[\l]$. Not only does it select
$[\l]$ uniquely in generic situations but it is also free from the
above drawbacks. Thus, on a generic WIH, there is a way to select
a canonical $[\l]$ \textit{using only the intrinsically available
structure} on $\IH$ such that, in the case of a Killing horizon,
$[\l ]$ consists precisely of the constant multiples of the
Killing vector.

On any  WIH, the flow generated by $[\l]$ preserves the metric and
also parts of the connection $\D$. These conditions are sufficient
for generalizing black hole mechanics \cite{abf,afk,abl2}.
However, from the geometric perspective of this paper, it is more
natural to impose a stronger requirement and demand that the flow
of $[\l ]$ preserve the full connection $\D$. This condition
defines the isolated horizons (IHs) of Section \ref{s5}. The
additional conditions are true restrictions in the sense that,
while every NEH can be made a WIH by selecting an appropriate null
normal, the same is \textit{not} true of IHs. Again, we analyze
the constraints imposed by the field equations, isolate the `free
data' and point out that, if a NEH geometry admits two IH
\textit{non-extremal} IH structures, they are related by a
(geometry preserving) diffeomorphism. Thus, if it exists, a
non-extremal IH structure is unique. In Section \ref{s6} we
consider the analytic extension of $\IH$ and its intrinsic
geometry and present a more geometric characterization of the
canonical normals $[\l]$. Section \ref{s7} summarizes our main
conclusions and suggests potential applications of our results.

In Appendix \ref{sa1} we obtain some explicit conditions that a
NEH geometry must satisfy if it is to admit an IH structure and if
it is to admit two distinct IH structures, one extremal and the
other non-extremal. Generically the IH structure is unique; the
exceptional cases are now even more restricted than they were in
Section \ref{s4.2}. Up to this point, we work with tensors in
index-free or Penrose's abstract index notation \cite{pr}. For the
convenience of readers more familiar with the Newman-Penrose
framework \cite{np1,pr}, in Appendix \ref{sa2} we derive the main
results as well as a few other interesting facts in that
framework.

\section{\bf Non-expanding horizons}
\label{s2}

In this section, we introduce the notion of non-expanding horizons
(NEH) and review some of their properties. This discussion sets
the stage for the main definitions introduced later in the paper.
Most of the properties summarized here were discussed in
\cite{abf}. They are included here for completeness and will be
presented from a more geometric perspective. For an extension of a
part of the framework to general, null hypersurfaces, see
\cite{td,jkc}.

\subsection{Preliminaries}
\label{s2.1}

Consider a 4-dimensional space-time $(M, g)$ and a 3-dimensional,
null sub-manifold $\IH$ thereof.  We will denote a(n arbitrary)
\textit{future-directed} null normal to $\IH$ by $\ell$.

\textbf{Definition 1:}  $\IH$ will be called a
\textit{non-expanding horizon} if \\
\noindent $i)$  $\IH$ is  diffeomorphic to the product
$\hIH \times \R$ where $\hIH$ is a 2-sphere, and the fibers of the
projection
$$ \Pi\ :\ \hIH \times \R \rightarrow \hIH $$
are null curves  in $\IH$; \\
\noindent $ii)$ the expansion of any null normal $\l$ to $\IH$
vanishes; and, \\
\noindent $iii)$ Einstein's equations hold on $\IH$ and the
stress-energy tensor $T_{ab}$ is such that $-T^a{}_b \l^b$ is
causal and future-directed on $\IH$.\\
\noindent Note that if these conditions hold for one choice of
null normal, they hold for all. Condition $iii)$ is very mild; in
particular, it is implied by the (much stronger) dominant energy
condition satisfied by the Klein-Gordon, Maxwell, dilaton,
Yang-Mills and Higgs fields as well as by perfect fluids.

Condition $i)$ above implies $\IH$ is ruled by the integral
curves of the null direction field which is normal to it. For
later purposes, it is useful to introduce an equivalence relation:
$$ \ell'\ \sim\ \ell \ {\rm whenever}\ \ \ell'\ =\ c\ell, \ \quad
{\rm where} \ \ c=\const. $$
and denote the equivalence classes by $[\ell ]$. Using the common
terminology at null infinity, the integral curves of $[\l]$ will
be called \textit{generators} of $\IH$ and $\hIH$ will be called
the \textit{base space} of $\IH$.  $i)$ also implies that the
generators of $\IH$ are geodesic.  Given $\ell$, a function
$$
v\ :\ \IH\ \rightarrow\ \R,\ \ {\rm such\ \ that}\ \quad \L_{\l}
\, v = 1
$$
will be called a {\it compatible coordinate}. Finally,
the Raychaudhuri equation and condition $iii)$ imply that $\l$ is
also \textit{shear-free} on $\IH$.  Hence, condition $ii)$ may be
replaced by:\\
\noindent{\it $ii)'$} The null direction tangent to $\IH$ is
covariantly constant on $\IH$.\\
Similarly, $i)$ and $ii)$ may be replaced by: \\
\noindent{\it $i,ii)''$} $\IH$ is isometric to the orthogonal
product of a 2-sphere $\hIH$ equipped with a positive definite
metric $\q_{ab}$ and the line $\R$ equipped with the trivial, zero
`metric'.\\
\noindent In Definition 1, we chose the formulation in terms of
notions that are most commonly used in the relativity literature.

Let us now examine the geometry of non-expanding horizons (NEHs).
First, via pull-back, the space-time metric $g_{ab}$ induces a
`metric' $q_{ab}$ on $\IH$ with signature (0,+,+).  Since $q$ is
degenerate, there exist infinitely many (torsion-free) derivative
operators on $\IH$ which annihilate it. However, the Raychaudhuri
equation and fact that $\IH$ is divergence-free and matter
satisfies our energy condition imply that $\IH$ is also
shear-free. We will now show that the vanishing of expansion and
shear in turn imply that we can uniquely select a preferred
derivative operator intrinsic to $\IH$. Let us first note that the
condition $(ii')$ implies the space-time parallel transport
restricted to the curves tangent to $\IH$ preserves the tangent
bundle $T(\IH)$ to $\IH$. Indeed, at each point $x\in \IH$, the
tangent space $T_x\IH$ is the subspace of $T_x M$ orthogonal to
$\ell$, and\relax
\footnote{Throughout this paper $\eqhat$ stands for ``equals, on
$\IH$, to"  and, unless otherwise stated, $d$ will denote the
exterior derivative intrinsic to $\IH$.}
$$
(X^a\nabla_a Y^b)\l_b \= -X^a Y^b \nabla_a \ell_b  \=0
$$
for every $X, Y\in T(\IH)$, where $\nabla$ is the derivative operator
on $M$ compatible with $g$. Therefore, $\nabla$ induces a derivative
operator $\D$ on $\IH$ via
\be X^a D_a Y^b = X^a\nabla_a Y^b \ee
for all vector fields $X,Y$ tangential to $\IH$. The operator $\D$
can be extended, in the standard fashion, to the covectors $K_a$
defined intrinsically on $\IH$: $ (\D_a K_b)Y^b \ = \D_a (K_bY^b)
- K_b (\D_a Y^b) $. Finally, the action of $\D$ can be uniquely
extended by the Leibnitz rule to arbitrary tensor fields defined
intrinsically on $\IH$. Pulling back the equation $\nabla_a g_{mn}
\= 0$ to $\IH$ we obtain $\D_a q_{mn} \=0$; $\D$ is compatible
with the degenerate metric on $\IH$.

By \textit{geometry of $\IH$ we will mean the pair $(q, \D)$
consisting of the intrinsic metric $q$ and the derivative $\D$ on
$\IH$}. Because $\ell$ is shear- and expansion-free, ${\cal
L}_{\ell}\, q \=0$ for any null normal $\ell$ to $\IH$.
Furthermore, by construction, $q_{ab} \ell^b \= 0$.  Hence we
conclude that $q_{ab}$ is the pull-back to $\IH$ of a Riemannian
metric $\q_{ab}$ on the 2-sphere $\hIH$ of generators of $\IH$:
$$ q \= \Pi^\star \,\, \q $$
Similarly, the natural (area) 2-form $\hat{\epsilon}$ compatible
with ${\q}$ on $\hIH$ can be pulled-back to yield a natural 2-form
$$ {}^{2}\!\epsilon \= \Pi^\star\,\, \hat\epsilon $$
on $\IH$, which will turn out to be useful.  Since $\q$ is
non-degenerate, it defines an unique (torsion-free) derivative
operator $\hat{\D}$ on $\hIH$.  If $h$ is the pull-back to $\IH$
of any covector $\hat{h}$ on $\hIH$, we have
\be \label{horD} \D_a\, h_b = \Pi^\star\,
(\hat{\D}_a\,\hat{h}_b)\ee
Therefore, in particular, we have:
$$ \D_a \, {}^{2}\!  \epsilon_{bc} \= 0\, .   $$

However, $\D$ is \textit{not} determined by $\hat{\D}$; it has
`more information'. In particular, $\hat{\D}$ does not constrain
the action of $\D$ on $\ell$ which determines `surface gravity'
and `gravitational angular momentum' of $\IH$ \cite{abl2}. Let
us extract this part of the extra information. Since
$X^aY^b\nabla_a \l_b \= 0$ for all $X,Y$ tangential to $\IH$,
there exists a 1-form $\omega$ on $\IH$ such that
\be \label{omegadef} \D_a\ell^b \= \omega_a \l^b \, . \ee
By construction, $\omega$ is tied to a choice of $\l$. (Strictly,
we should denote it as $\omega_{(\ell)}$ but will refrain from
doing so for notational simplicity.) Under the rescaling $\ell
\mapsto f \ell$, we have
\be \label{omegatrans} \omega \mapsto \omega + d \ln f \, . \ee
The function $\kl$ defined by
$$ \kl := \omega_a \l^a $$
will be called \textit{surface gravity} of $\IH$ relative to the
null normal $\ell$.  Under the rescaling $\ell \mapsto f \ell$, we
have
\be \label{trans1} \kl \mapsto \kl + {\cal L}_\ell \ln f . \ee
Hence, given only a NEH, it is not meaningful to ask if the
surface gravity is constant on $\Delta$, i.e., if the zeroth law
of black hole mechanics holds. It can hold for one choice of
$\ell$ but not for another. This brings out the fact that the
notion of a non-expanding horizon does not capture even the most
basic \textit{physical} structure available on the event horizon
of a black hole in equilibrium.  In Section \ref{s3} we will
strengthen the definition by adding suitable conditions to ensure
that the zeroth law does hold.

\subsection{Space-time curvature on $\IH$}
\label{s2.2}

The geometry $(q,\D)$ of $\IH$ determines the pull-back
$R^a{}_{\pback{bcd}}$ (on the three covariant indices) of the
space-time Riemann tensor $R^a{}_{bcd}$, and, by algebraic
symmetries of $R^a{}_{bcd}$, the pull-back $R_{\pback{ab}}$ of the
Ricci tensor $R_{ab}$. Our energy condition (in Definition 1) and
the Raychaudhuri equation on $\l$ not only force the shear of $\l$
to vanish, but also imply the Ricci tensor must satisfy:
\be \label{riccill} R_{ab} \ell^a \l^b \=0. \ee
Our energy condition then further implies that $R_{ab}\ell^b$ is
proportional to $\ell_a$, that is
\be\label{riccil} R_{ab}\ell^a X^b\ \=\ 0, \ee
for any vector field $X$ tangent to $\IH$. In the Newman-Penrose
notation (see Appendix \ref{sa2}), this result can be stated as
follows: in a null frame $(m,\,\bar{m},\, n,\, \ell)$,
\be \label{ricci} \Phi_{00}\ \=\ \Phi_{01}\ =\bar\Phi_{10}  \= 0.
\ee
(This equation, in turn, constrains the matter fields via
$T_{ab}\l^a X^b \= 0$. However, in what follows we are primarily
interested in the geometrical fields.) The remaining components of
$R_{\pback{ab}}$ enter `constraint equations' on $\IH$ and are
discussed in Section \ref{s3.2}.

Let us next consider the Weyl tensor. Because $\ell$ is expansion
and shear-free, it must lie along one of the principal null
directions of the Weyl tensor (see Appendix \ref{sa2} or
\cite{abf}). Then, equation (\ref{riccil}) implies $\ell$ in fact
lies along a \textit{double} principal null direction of the Weyl
tensor \cite{abf}. In the Newman-Penrose notation we have:
\be \label{weyl}  \Psi_0\ \=\ \Psi_1 \= 0,  \ee
whence $\Psi_2$ is gauge invariant (i.e., independent of the
choice of the null-tetrad vectors $(n,m,\bar{m})$) on $\IH$. The
remaining components of $R^a{}_{\pback{bcd}}$ will not be needed
in the main text but are given at the end of the subsection 2.a of
the Appendix \ref{sa2} for completeness.

Next, let us explore the relation between the intrinsic derivative
operator $\D$ on $\IH$ and the (non-vanishing components of the)
Weyl curvature.  Note first that while the 1-form $\omega$ of
(\ref{omegadef}) depends on the choice of $\ell$, it is clear from
its transformation property (\ref{omegatrans}) that its exterior
derivative $d\omega$ is in fact independent of $\ell$.
Furthermore, a simple calculation \cite{afk} shows $d\omega$
can be expressed in terms of the Riemann curvature. Using
(\ref{omegadef}) and (\ref{riccil}) we have
\be \label{domega} d\omega \= \ 2 \Im(\Psi_2)\,\, {}^2\epsilon. \ee
As one's experience with the Newman Penrose framework would
suggest, $\Im(\Psi_2)$ captures the gravitational contribution to
the angular-momentum at $\IH$ \cite{abl2}. Therefore, we will
refer to $\Im(\Psi_2)$ as the {\it rotational curvature scalar} and
$\omega$ as the {\it rotational 1-form potential}.

Using the Cartan identity and (\ref{domega}), the Lie derivative
of $\omega$ with respect to $\ell$ is given by
\be\label{dkappa} \L_\ell \omega_a \=\ 2\Im(\Psi_2)\, \l^b\,
{}^2\!\epsilon_{ba} + \D_a (\l^b\omega_b)\ \=\ \D_a\kl. \ee
Recall that $\kl$ is associated with a pair $(\IH, \ell)$ rather
than with the 3-manifold $\IH$ itself and it changes under
rescalings of $\ell$ via (\ref{trans1}). Eq (\ref{dkappa})
provides a necessary and sufficient condition on the choice of
$\ell$ to ensure that $\kl$ is constant on $\IH$, i.e., the zeroth
law holds. It will motivate our definition of weakly isolated
horizons in the next section.

Finally, we will show that the rotational curvature scalar $\Im
(\Psi_2)$ also admits a \textit{scalar} potential, which turns out
to be useful. Note first that (\ref{domega}) and (\ref{dkappa})
imply $\Im(\Psi_2)$ has an unambiguous projection to $\hIH$:
$$
{\cal L}_\ell \,\, \Im(\Psi_2)\ \= 0,
$$
and, considered as a function defined on $\hIH$, it satisfies
a `global constraint', namely
\be
\int_{\hIH}\,  2\Im(\Psi_2) \, \hat\epsilon \ =\ 0.
\ee
Therefore, there exists a well-defined {\it rotational scalar
potential} $U$ such that
\be \label{scalarpot} \hDelta U\ =\ 2\Im(\Psi_2), \ee
on $\hIH$ which is unique up to an additive constant: the only
freedom in the choice of $U$ is

$$U\mapsto  U'\ =\ U + U_0,\ \ U_0\ =\ \const. $$

\section{Weakly isolated horizons $[\l ]$}
\label{s3}

As noted above, on a NEH there is a large freedom in the choice of
null normals corresponding to changes  $\l \mapsto \l^\prime \=
f\l$ with $f$ arbitrary positive function on $\IH$. Applications
of this framework, e.g. to black hole mechanics and numerical
relativity, often require the horizon to be equipped with a
preferred equivalence class $[\l ]$ \cite{prl,abf,afk,abl2}.
Therefore, in this section, we will endow NEHs with a specific
$[\l]$ satisfying a weak restriction, which enables one to extend
zeroth and first laws of black hole mechanics, and study the
resulting geometrical structures.

This section is divided into three parts. In the first, we
introduce the basic definition of a weakly isolated horizon
(WIH); in the second, we examine the interplay between the
space-time geometry and the intrinsic structures on WIHs; and, in
the third, we show that (non-extremal) WIHs admit a preferred
foliation. These structures are useful not only for mathematical
physics but also for numerical relativity \cite{prl}.

\subsection{Preliminaries}
\label{s3.1}

\noindent{\bf Definition.} A \textit{weakly isolated horizon}
$(\IH, [\ell])$ is  a non-expanding horizon $\IH$, equipped with
an equivalence class $[\ell]$ of null normals (under
\textit{constant} rescaling) such that the flow of $\ell$
preserves the rotation 1-form potential $\omega$ (of
(\ref{omegadef}))
\be\label{wih} \L_\ell \, \omega_a \ \=\ 0. \ee
If this condition holds for one $\l$, it holds for all $\l$ in
$[\l ]$.

Note that, by definition, a WIH is equipped with a specific
equivalence class $[\l]$ of null normals. Recall that, on a
space-like hypersurface $S$, the extrinsic curvature can be
defined as $K_{a}{}^b = \nabla_{\!\pback{a}} n^b$ where $n$ is the
unit normal and under-arrow denotes the pull-back to $S$. A
natural analog of the extrinsic curvature on a WIH is then
$L_a{}^b \= \D_a \l^b$ and, by virtue of (\ref{omegadef}),
(\ref{wih}) is equivalent to requiring that $L_a{}^b$ be
Lie-dragged along $\l$.\relax
\footnote{Of course, on a null surface there is no canonical
analog of extrinsic curvature. For example, since the pull-back of
$\l_a$ to $\IH$ vanishes, $L_{ab} := \nabla_{\pback{a}} \l_b$
vanishes identically. Thus, the index structure of $L_a{}^b$ has
to be chosen carefully if one wishes to strengthen the notion of a
NEH.}
Thus, while on a NEH only the intrinsic metric $q$ is
``time-independent'', on a WIH the analog of extrinsic curvature
is also required to be ``time-independent''. In this sense, while
NEHs resemble Killing horizons only to first order, WIH resemble a
Killing horizon also to the next order. Note however that the full
connection $\D$ or curvature components such as $\Psi_4$
\textit{can be} time-dependent on a WIH.

Eqs (\ref{wih}) and (\ref{dkappa}) imply $\kl$ is constant on
$\IH$.  Thus \textit{the zeroth law of black hole mechanics
naturally extends to} WIHs. However, since a WIH is equipped only
with an equivalence class $[\ell ]$ of null normals, where $\ell
\sim \ell^\prime$ if and only if $\ell^\prime = c \ell$ with $c$
a positive constant, and since $\kappa_{(l^\prime)}= c \kl$ by
(\ref{trans1}), that surface gravity does not have a
canonical value on WIHs unless it vanishes. Thus, WIHs naturally
fall in to two classes: i) non-extremal, in which case the surface
gravity for every $\ell$ in $[\ell ]$ is non-zero; and, ii)
extremal, in which case it is zero.

Given any NEH  $\IH$, we can always choose a null normal $[\l ]$
such that $(\IH, [\l ])$ is an extremal WIH. In this case, $\ell^a
\nabla_a \ell^b \= 0$; integral curves of $\ell \in [\ell]$ are
\textit{affinely parametrized geodesics}. Fix an $\l_0$ which is
affinely parameterized on $\IH$ and denote by $v_0$ a compatible
coordinate (so ${\cal L}_{\ell_0} \, v_0 \= 1$). On the same
manifold $\IH$, consider any other null normal $\ell^\prime_0$
such that $(\IH, [\ell^\prime_0])$ is also an extremal WIH. Let
$v^\prime_0$ be a compatible coordinate for $\ell^\prime_0$. Then,
it is straightforward to check that
\be \label{aff}
 \ell^{\prime a} \= (\frac{1}{A}) \ell^a  \quad {\rm and}\quad
   v^\prime_0 \= Av_0 +B
\ee
for some  functions $A$ and $B$ on $\IH$ such that ${\cal L}_\ell
A \= {\cal L}_\ell B \= 0$ and $\quad A > 0.$
Thus, every NEH admits a family of null normals $\ell,
\ell^\prime, ....$ such that $(\IH, [\ell ])$, $(\IH, [\ell^\prime
])$, .... are all extremal WIHs and any two of these null normals
are related by (\ref{aff}).

Next, let us examine the rescaling of $\l_0$ which maps the
fiducial, extremal WIH to any given, non-extremal WIH  $(\IH,
[\l])$ with surface gravity $\kl$ (see (\ref{trans1})). It is
given by:
\be \label{WIHall} \ell^a \= \kl\, (v_0 - B)\, \l_0^a \quad {\rm
and}\quad  \kl v =  \ln\, \kl (v_0 -B)\ee
where $\L_{\l} B \= 0$. Thus, every non-extremal WIH $(\IH, [\ell
])$ is obtained from the fiducial $\ell_0$ via (\ref{WIHall}). To
summarize, simply by restricting the null normals $\l$ to lie in a
suitably chosen equivalence class $[\l]$, from any given NEH $\IH$
we can construct a WIH $(\IH, [\l])$ which is either extremal or
non-extremal. However, because of the arbitrary functions involved
in (\ref{aff}) and (\ref{WIHall}), \textit{there is an infinite
dimensional freedom in this construction.}\\

We will conclude this sub-section with four remarks.

i) We could also begin with a non-extremal WIH and construct an
extremal one.  Let $\ell$ be a null normal to $\IH$ such that $\kl
= \const \not\= 0$.  Fix any function $v$ on $\IH$ with $\L_{\ell}
v \= 1$. Then, if we set $\ell^\prime \= (\exp\ {-\kl v}) B\,\ell
$ where $B$ is any function on $\IH$ satisfying ${\cal L}_\ell B
\= 0$, then $(\IH, [\ell^\prime ])$ is an extremal WIH and every
extremal WIH can be obtained via this construction.

ii) Note that in (\ref{WIHall}), $\ell$ vanishes if $v_0 = B$.
Since in the main text, for simplicity, we have restricted
ourselves only to future directed, non-zero null normals, strictly
speaking, the construction breaks down and only a portion of
$(\IH, [\ell_0 ])$ can be regarded as a non-extremal WIH. However,
geometrically, there is no a priori obstruction to allow $\l$ to
vanish somewhere and consider the entire solution (\ref{WIHall}).
Then $\ell$ changes orientation at points $v_0 = B$.  In such
situations, the section of $\IH$ defined by $\ell=0$ will be
called the {\it crossover section of $\ell$} (which may not have a
2-sphere topology if $\l_0$ fails to be complete on $\IH$). Notice
however that from the perspective of the geometry of the NEH
$\IH$, there is nothing special about the crossover section of a
given WIH. Indeed, \textit{every} section of $\IH$ is a crossover
section of some non-extremal WIH. In Section \ref{s6}, we will
consider the geodesically complete, analytic extension $\bar\IH$
of $\IH$. In such an extension, a non-extremal WIH always contains
a \textit{2-sphere} cross-section on which $\l^a$ vanishes. It
will be called a {\it cross-over sphere.} In the Kruskal extension
of the Schwarzschild space-time, the Killing horizon bifurcates at
the cross-over sphere.

iii) Suppose a WIH $(\IH, [\ell ])$ is complete in the sense that
each $\ell \in [\ell ]$ is a (future directed, nowhere vanishing)
complete vector field. Then, given a representative $\ell\in
[\ell]$, the corresponding rotation 1-form $\omega$ may be thought
of as an Abelian connection on the trivial bundle $\IH\rightarrow
\hIH$, where the structure group is (the additive group of reals)
given by the flow generated by $\ell$.

iv) Let $\IH_K$ be a Killing horizon for Killing fields $[\xi ]$,
where, as before, the square brackets denote  equivalence class of
vector fields where any two are related through rescaling by a
positive constant. If we set $[\l ] = [\xi ]$, then $(\IH, [\l ])$
is a WIH. Thus, the passage from the NEH to a WIH can be
understood as follows: whereas on a NEH we only ask that the null
normal be a Killing field to the first order (i.e., it Lie drag
the intrinsic metric $q_{ab}$ on $\IH$), on a WIH, the permissible
null normals mimic the Killing fields in a stronger sense; they
Lie drag also the connection 1-form $\omega$.

\subsection{Constraint equations and free data on WIHs }
\label{s3.2}

On a space-like 3-manifold, the 4-geometry induces an intrinsic
metric and an extrinsic curvature and these are subject to the
well-known constraint equations. Under the weak assumption that
space-time admits constant mean curvature slices, one can find
the freely specifiable data through the Lichnerowicz-York
construction \cite{jy}. Similarly, at null infinity, $\I$, the
(conformally rescaled) 4-metric naturally induces a triplet $(q,
n, \D )$ consisting of a degenerate metric $q$, null normal $n^a$
and an intrinsic derivative operator $\D$ \cite{aa}. These fields
capture the information contained in the pull-back of the 4-Ricci
tensor to $\I$, and five of the ten components, $\Psi_4, \Psi_3$
and $\Im\, \Psi_2$, of the Weyl tensor. The constraint equations
they satisfy enable one to isolate the radiative degrees of
freedom of the gravitational field. We will now carry out a
similar analysis at weakly isolated horizons. More precisely, we
ask: what are the constraints which the triplet $(q_{ab}, [\l^a],
\D)$ must satisfy and can they be solved to obtain the ``freely
specifiable data'' at WIHs?

The definition of a WIH immediately leads to the first set of
equations:
\ba \label{kin} q_{ab} l^b & \= & 0,\qquad \L_{\l}\, q_{ab}  \= 0
\nonumber\\
\D_a \l^b & \= & \omega_a \l^b , \quad \L_{\l} \omega_a \=  0. \ea
Since $q$ is degenerate, it does not fully determine $\D$.
Nonetheless, as noted in (\ref{horD}), it does constrain $\D$: if
$h$ is the pull-back to $\IH$ of a 1-form field $\hat{h}$ on
$\hIH$, then
$$ \D_a h_b = \Pi^* (\hat{D}_a \hat{h}_b) $$
where $\hat{\D}$ is the unique (torsion-free) derivative operator
on $\hIH$ compatible with $\hat{q}$. Therefore, given $q$, to
specify the action of $\D$ on an arbitrary co-vector field $W$ on
$\IH$ ---and hence on any tensor field on $\IH$--- it is necessary
and sufficient to specify its action on a co-vector field $n$ with
$n_a\l^a \not\= 0$. Thus, we only need to specify
$$\S_{ab} := \D_a n_b \, .$$
Without loss of generality, we can choose $n$ to satisfy
\be \label{n}
 n_a\l^a \= -1; \qquad {\rm and} \qquad
 \D_{[a} n_{b]} \= 0\, . \ee
(Note that (\ref{n}) is equivalent to setting $n = dv$, with $v$ a
compatible coordinate for $\ell$.) These properties (\ref{n}) now
imply that $\S_{ab}$ is symmetric and satisfies
\be \label{Nl}  \S_{ab}\l^b \=  \omega_a \ee
Hence, given $(q, \omega)$, to specify $\D$, it suffices to
provide just the projection $\tilde{\S}_{ab}$ of $\S_{ab}$ on
2-sphere cross-sections of $\IH$ orthogonal to $n$:
$$ \tilde{\S}_{ab} \= \tilde{q}_a{}^c \tilde{q}_b{}^d \S_{cd} $$
where $\tilde{q}_a{}^c$ is the projection operator on these
2-spheres (satisfying $\tilde{q}_a{}^b \l^a \= 0 \=
\tilde{q}_a{}^b n_b$, and $\tilde{q}_a{}^b \tilde{X}^a =
\tilde{X}^b$ for all $\tilde{X}^b$ tangential to the 2-sphere
cross-sections.) The trace $\tilde{q}^{ab} \tilde{\S}_{ab}=: 2\mu
$ of $\tilde\S_{ab}$ represents the ``transversal'' expansion of
the 2-spheres while its trace-free part $\lambda_{ab}$ represents
their ``transversal'' shear (where ``transversal'' refers to $n$).
Thus, we have shown that the geometry $(q,\D)$ of $(\IH, [\l])$ is
completely specified by $(q, \omega, \tilde{\S}_{ab})$. The
question now is: What are the restrictions imposed on this triplet
by the field equations $R_{ab} = 8\pi G (T_{ab} - \frac{1}{2} T
g_{ab})$ at $\IH$?

It turns out that the components of $R_{ab}$ transversal to $\IH$
dictate the ``evolution'' of fields off $\IH$ while the pull-back of
$R_{ab}$ constrains its intrinsic geometry $(q, \D)$. A direct
calculation yields: $R_{\pback{ab}}\l^b \= 2 \l^a \D_{[a}
\omega_{b]}$. But we have already seen that $R_{\pback{ab}}\l^b$
vanishes identically on every NEH (see (\ref{riccil}). Using
(\ref{wih}) we conclude that this vanishing implies and is implied by
the zeroth law, i.e., the condition that $\kl =\omega_a\l^a$ is
constant on $\IH$.  (The field equations also constrain matter fields
in the obvious way but these restrictions are not relevant for
$(q,\omega, \tilde{\S})$.) Next, another direct calculation yields
\be \label{ls1} \L_\l \, \tilde{\S}_{ab} \= -\kl \tilde{\S}_{ab} +
\tilde\D_{(a} \tilde\omega_{b)} + \tilde\omega_a \tilde\omega_b -
\frac{1}{2} \tilde{\cal R}_{ab} + \frac{1}{2} \tilde{q}_a{}^c\,
\tilde{q}_b{}^d \, R_{cd} \ee
where $\tilde{\D}$ and $\tilde{\cal R}_{ab}$ denote the derivative
operator and the Ricci tensor on the 2-sphere cross-sections
defined by $n_a$ and $\tilde \omega_a$ is the projection of
$\omega_a$ on these cross-sections.  Thus, by (\ref{riccil}) the
contraction of the pulled-back Ricci tensor with $\l$ vanishes,
while by (\ref{ls1}) its remaining components serve as the source
of the time derivative of $\tilde{\S}_{ab}$. Hence, the constraint
equations on $(q,\D )\, \equiv \, (q, \omega, \tilde{\S})$ are
simply the zeroth law the
\be \label{1law} \kl \= {\rm const}\ee
and the equation
\be \label{ls2} {\cal L}_\l \, \tilde{\S}_{ab} = -\kl
\tilde{\S}_{ab} + \tilde\D_{(a} \tilde\omega_{b)} + \tilde\omega_a
\tilde\omega_b - \frac{1}{2} \tilde{\cal R}_{ab} + 4\pi G
\tilde{q}_a{}^c\, \tilde{q}_b{}^d \, (T_{cd} - \frac{1}{2} T
q_{cd})\, , \ee
where we have used the field equations in the last term. Having
these constraints at our disposal, we can now spell out the freely
specifiable data.

Suppose we are given a pair, $(\IH, [\l ])$, (satisfying condition
$i)$ of Definition 1) and the pull-back and the trace of the
stress-energy tensor $T_{ab}$ on $\IH$. To construct the
permissible pairs $(q_{ab}, \D)$ such that $(\IH, [\l ])$ is a
weakly isolated horizon, we proceed as follows. Choose any
2-sphere cross section $\tilde{\IH}$ of $\IH$ and fix a Riemannian
metric $\tilde q_{ab}$, a 1-form $\tilde\omega_a$, and a symmetric
tensor $\tilde\S_{ab}$ on $\tilde\IH$.  Extend these fields to all
of  $\IH$ in two stages: First, set $\ell^a \tilde q_{ab} = 0 =
\ell^a \tilde S_{ab}$ and $\ell^a \omega_a = \kl$ at $\tilde\IH$,
with $\kl$ an arbitrary positive constant.  Second, carry these
fields to other points $\IH$ by setting ${\cal L}_\ell \tilde
q_{ab} = 0 = {\cal L}_\ell \tilde \omega_a$ and using (\ref{ls2}).
Then, we have a permissible pair $(q = \tilde q, \D)$, where $\D$
is constructed from $\tilde q$, $\tilde \omega$ and $\tilde S$.
All permissible pairs can be constructed in this way. Thus, given
$(\IH, [\l])$ the free data consist of a constant $\kl$ and fields
$\tilde{q}_{ab}, \tilde{\omega}_a, \tilde{\S}_{ab}$ on any
2-sphere cross-section of $\IH$. In the Newman-Penrose notation,
$\tilde\omega_a$ is specified by the spin coefficient $\pi$ and
$\tilde{\S}_{ab}$ by the coefficients $\mu$ and $\lambda$.

We conclude this sub-section with four remarks.

i) In the standard 3+1 decomposition of space-time by spatial
sub-manifolds, the pull-back of the space-time Ricci tensor to the
sub-manifolds yields the evolution equations along the normal to
the surface. In the present case, since the normal $\l$ is also
tangential to $\IH$, the analogous equation (\ref{ls1}) is now a
constraint.

ii) At null infinity, $\I$,  because of the available conformal
freedom, we can choose the analog of $q$ to be the  standard
2-sphere metric, and the analog of $\omega$ as well as the trace
of the analog of $\tilde{\S}_{ab}$ to be zero. The (conformal
equivalence class) of the derivative operator $\D$ at
$\mathcal{I}$ is thus fully determined by the transversal shear.
In absence of radiation, we again have an analog of (\ref{ls2}),
but it now says that the transversal shear is
\textit{time-independent}. Thus, while the overall mathematical
structure is parallel, the detailed conclusions are different
because of differences in the boundary conditions in the two
regimes.

iii) Since the above construction involved the choice of a
cross-section $\tilde{\IH}$ of $\IH$, there is a certain gauge
freedom in the free data. Suppose we determine the geometry of
$\IH$ by choosing a $\tilde\IH$ and carrying out the construction
to obtain a triplet $(q, \omega, \S)$. Then, if we choose a new
cross-section $\tilde{\IH}^\prime$ and use as initial data the
fields induced on $\tilde{\IH}^\prime$ by $(q,\omega, \S)$, our
construction will yield new fields $(q^\prime, \omega^\prime,
\S^\prime)$ on $\IH$, gauge related to the original triplet.  The
gauge transformations are:
$$ q^\prime_{ab} = q_{ab}, \quad \omega^\prime_a = \omega_a,
\quad n^\prime_a = n_a+ \D_a f, \quad \S_{ab}^\prime = \S_{ab} +
\D_a\D_b f
$$
for some function $f$ satisfying $\L_\l f \=0$. Hence,
$\tilde\omega$ and $\tilde{\S}_{ab}$ transform via
\be \label{trans2} \tilde\omega^\prime_a = \tilde\omega_a -\kl
\D_a f \quad {\rm and} \quad \tilde{\S}^\prime_{ab} =
\tilde{\S}_{ab} +  \D_a \D_b f\, . \ee
In spite of this non-trivial transformation property, $\L_\l
\tilde{\S}_{ab} \= \L_\l\tilde{\S}^\prime_{ab}$. This fact will
play an important role in the next two sections. Finally, we will
show in Section \ref{s3.3} that on non-extremal WIHs this gauge
freedom can be completely eliminated by a natural choice of
cross-sections.

iv) Let us suppose the pull-back of the space-time Ricci tensor to
$\IH$ is time-independent, i.e., $\L_\l R_{\pback{ab}} \= 0$.
(Incidentally, because of (\ref{riccil}), if this property holds
for one null normal to $\IH$, it holds for all.) Then all but the
first term on the right side of (\ref{ls1}) are ``time
independent'' whence, we can easily integrate this equation. If
$\kl$ is non-zero, the solution is:
\be\label{sol1}  \tilde\S_{ab} = e^{{-\kl}v} \, \tilde{S}^0_{ab} +
\frac{1}{\kl}\left[\tilde\D_{(a} \tilde\omega_{b)} + \tilde\omega_a
\tilde\omega_b - \frac{1}{2} \tilde{\cal R}_{ab} + \frac{1}{2}
\tilde{q}_a{}^c\, \tilde{q}_b{}^d \, R_{cd}\right] \ee
for some $v$-independent $\tilde{S}^0_{ab}$, while if $\kl$
vanishes, it is:
\be\label{sol2}  \tilde\S_{ab} =  \tilde{S}^0_{ab} +
\left[\tilde\D_{(a} \tilde\omega_{b)} + \tilde\omega_a \tilde\omega_b -
\frac{1}{2} \tilde{\cal R}_{ab} + \frac{1}{2} \tilde{q}_a{}^c\,
\tilde{q}_b{}^d \, R_{cd}\right] v \ee
Thus, in either case, $\tilde\S_{ab}$ has a very specific time
dependence\relax
\footnote{The possibility of time dependence of these fields
brings out the generality of the notion of WIHs. On a Killing
horizon, by contrast, every geometrical field is time
independent.}\relax
.  Let us suppose that the WIH $(\IH, \l)$ is complete and
non-extremal. Then $\tilde{\S}_{ab}$ diverges at one end unless
$\tilde\S^0_{ab}$ vanishes identically. By remark iii) above, this
property is independent of the choice of $n$. However, it does
depend on the choice of $\l$. In Section \ref{s4.2} we will use
this fact to select a canonical $[\l ]$ on a generic NEH $\IH$.

\subsection{Good cuts of a non-extremal WIH $(\IH, [\ell ])$}
\label{s3.3}

We will now show that non-extremal WIHs admit a natural foliation
which can be regarded as providing a `rest frame' for the horizon.
Fix a WIH with $\kl \not\= 0$. Since by the definition of a WIH
$\L_\ell d\omega \=0$, and since (\ref{domega}) implies that the
contraction of  $\ell$ with  $d\omega$ vanishes, $d\omega$ has an
unambiguous projection to the 2-sphere $\hIH$ of generators of
$\IH$. Since this projection is necessarily an exact 2-form, there
exists a 1-form $\homega$ defined globally on $\hIH$ such that
$$ \Pi^\star \hat{d}\homega\ =\ d\omega. $$
While $\homega$ is \textit{not} unique, \textit{any} choice will
give $d(\omega - \Pi^\star \homega) \= 0$ and $\ell^a (\omega_a -
\Pi^\star \homega_a) = \ell^a \omega_a = \kl$. Thus, there exists
a function $v$, defined globally on $\IH$, such that
\be\label{omegahat} \omega\ =\ \Pi^\star \hat\omega + \kl dv .\ee
Clearly, $v$ is determined up to an additive constant and is a
coordinate compatible with $\l$ (i.e., $\L_\l v = 1$). Hence, $v=
\const$ surfaces define a foliation of $\IH$. (In the extremal
case, $\omega = \Pi^\star \homega$ and we can no longer extract
the function $v$ or the foliation.) Thus, each choice of $\homega$
provides a foliation and the issue is if there is a natural,
invariant choice.

The answer is in the affirmative. Using the 2-metric $\q_{ab}$ and
the Hodge star $\hat{\star}$ on $\hIH$, any 1-form $\homega$ can
be uniquely decomposed in to exact and a co-exact parts:
$$ \label{dec}
\homega\ =\ - \hat{\star} \hat{d}U \ +\ \hat{d}p, $$
where $U,p$ are smooth, real functions on $\hIH$. $U$ is the
rotational scalar potential defined in Section \ref{s2.2}, while
$p$ represents the gauge-freedom in the choice of $\homega$. It is
natural to set it to zero, so that
\be\label{divfree} \homega\ =\ - \hat{\star}\hat{d}U. \ee
This prescription determines $\homega$ uniquely. Furthermore, this
is a natural choice because, e.g., when the rotational curvature
scalar $\Im(\Psi_2)$ vanishes, so does $\homega$ in this gauge. The
corresponding  $v=\const$ sections will be called
\textit{good cuts}. This foliation \textit{always} exists on
non-extremal WIHs and is invariantly defined in the sense that it
can be constructed entirely from structures naturally available on
$(\IH, [\l ])$. In particular, if the space-time admits an
isometry which preserves the given WIH, good cuts are necessarily
mapped in to each other by that isometry.\\

We conclude with two remarks.

i) In the definition of a WIH, we have not required
the vector fields $[\l]$ to be complete. This is because, in the
physically interesting situations, one expects the isolated
horizon to be formed at some finite time by a dynamical process
and hence incomplete at least in the past. When $[\l]$ is
incomplete, the good cuts defined above need not be global
cross-sections of $\IH$. However the requirement that $\homega$
is of the form (\ref{divfree}) globally on $\hIH$ still
distinguishes the good cut foliation and the associated
coordinates $v$ uniquely.

ii) At null infinity, in absence of radiative modes, it is also possible
to single out ``good cuts'' \cite{aa}. On WIHs, by definition,
there is no gravitational radiation. Is there then a close
relation between the two constructions? At first sight, the answer
may seem to be in the negative because, whereas the emphasis at
$\I$ is generally on finding cross sections on which the
transversal shear vanishes \cite{np2}, we made no reference to the
shear of $n$. (Indeed, as the transformation property
(\ref{trans2}) shows, the trace-free part of $\tilde{\S}_{ab}$ can
not be made to vanish on a general WIH.) Therefore, the two
constructions may appear to be unrelated. However, this is not the
case. In both cases, the absence of gravitational radiative modes
on the null 3-surface imposes a restriction on $\Im(\Psi_2)$,
enabling one to impose natural restrictions on its potential to
select good cuts. At null infinity, the potential happens to be
the transversal shear while in our case it happens to be the
1-form $\omega$. Thus, when formulated in terms of curvatures and
connections intrinsic to the null surfaces \cite{aa}, the two
constructions are parallel. There is nonetheless one residual
difference. Whereas the intrinsic degenerate metric on $\I$  has
considerable conformal freedom, that on $\IH$ is unique. Because
of this difference in `rigidity', whereas (in absence of radiative
modes) there is a 4-parameter family of good-cuts on $\I$, $\IH$
is equipped with a 1-parameter family.

\section{Preferred classes $[\l]$ of null normals}
\label{s4}

In applications of this framework, e.g. to numerical relativity,
one can find NEHs but it is generally essential to single out a
canonical equivalence class of null-normals $[\l]$. In the case
when the NEH arises as a Killing horizon $\IH_K$, the Killing
vector $\xi$ which is normal to $\IH_ K$ provides a canonical $[\l
]= [\xi ]$. (Note that, without recourse to global considerations,
$\xi$ is known only up to a multiplicative constant.) This choice
turns out to be the appropriate one for many applications. The
question is if we can select a canonical $[\l]$ on a generic NEH
such that, in the case of a Killing horizon, the canonical $[\l ]$
agrees with $[\xi ]$. In Section \ref{s4.1}, we consider an
`obvious' restriction which suffices to select $[\l ]$ uniquely.
While this procedure is rather natural from the standpoint of null
geometry, it turns out not to be satisfactory for various
applications. In Section \ref{s4.2}, therefore, we strengthen the
notion of `isolation' slightly by requiring that $q^{ab}R_{ab} \=
16\pi G T_{ab}\l^a n^b$ be time-independent on $\IH$ and introduce
a more sophisticated strategy which is also natural and
appropriate for applications.\relax
\footnote{Here $q^{ab}$ is any `inverse' of $q_{ab}$, i.e., any
tensor field defined intrinsically on $\IH$ satisfying $q^{ab}
q_{am} q_{bn} \eqhat q_{mn}$. There is a freedom to add to
$q^{ab}$ a term $X^{(a} \l^{b)}$, and to $n^a$ a vector field
$h^a$, where $X^a$ is \textit{any} vector field tangential to
$\IH$ and $h^a$ is any vector field tangential to $\IH$ satisfying
$\l \cdot h \eqhat -1$. However, our additional requirement is
insensitive to this freedom because $R_{ab}l^b X^b$ vanishes on
any $\IH$.}

\subsection{Canonical choice of $[\l ]$ on an extremal WIH}
\label{s4.1}

From the perspective of null surfaces, it is natural to limit the
freedom in the choice of $\l$ by first demanding that it be
\textit{affinely parametrized}. Then, $\kl$ vanishes, whence
$(\IH, [\l])$ is an extremal horizon. However, as we saw in
Section \ref{s3.1}, there is still an infinite dimensional freedom
in the choice of such $[\l]$. We will now show that this remaining
freedom can be eliminated by imposing a natural geometric
condition.

Let $\IH$ be a NEH and let $\l$ be a null normal such that $\kl
\= 0$. Since $\L_\l \omega_a$ and $\omega_a\l^a$ both vanish on
$\IH$, we have
$$ \omega\ =\ \Pi^* \homega, $$
whence the 1-form $\homega$ on $\hIH$ is defined uniquely by
$\omega$ on $\IH$. (In this case, the function $p$ in the
decomposition (\ref{omegahat}) no longer represents a gauge
freedom; in fact, $\hat{d}p$ is now an invariant of the WIH under
consideration.) Now, under a rescaling $\l \mapsto \l' \= C \l$ by
a function $C$, with $\L_\l C \= 0$, we have:
\be \omega'\ \= \ \omega + d\ln C,\quad \ p' = p + \ln C. \ee
Therefore, to select $\l$ uniquely up to a multiplicative
constant, it is necessary and sufficient to impose a condition
which selects $p$ up to an additive constant. Following our
strategy of Section \ref{s3.2}, we can achieve this by requiring
that $\hat\omega$ be divergence free on $\hIH$. To summarize,
\textit{any NEH $\IH$ admits an unique, extremal WIH structure
$(\IH, [\l])$ such that:}
\be \kl\ =0, \quad \hat{d}\, \hat{\star}\hat\omega = \hat{\div}
\homega\ \=\ 0 \ee
Purely from geometric considerations intrinsic to $\IH$, then, we
already have a prescription to select the equivalence class $[\l]$
of null normals uniquely.

Unfortunately, this prescription is not very useful in practice.
Suppose $\IH$ is a Killing horizon for a Killing vector $\xi$
defined in its neighborhood such that $\kappa_{(\xi)}$ is
\textit{non-zero}, i.e., the Killing horizon is non-extremal.
(This would, in particular, be the case for the Schwarzschild
horizon.) Then, not only will the unique equivalence class $[\l ]$
given by the above construction fail to coincide with the natural
choice $[\xi ]$ but more importantly, because of (\ref{WIHall}),
$[\l ]$ would not even be left invariant under the action of the
isometry generated by $\xi $. Indeed, from the 4-dimensional
space-time point of view, the assignment of \textit{any} extremal
$[\l ]$ to this Killing horizon would be unnatural. We therefore
need a more sophisticated strategy to single out a canonical
$[\l]$ on a NEH $\IH$. In particular, this choice should be left
invariant by all isometries preserving $\IH$.

\subsection{Canonical choice of $[\l ]$ on generic NEHs}
\label{s4.2}

Let us now restrict ourselves to NEHs $\IH$ such that the
space-time Ricci tensor satisfies $(\L_\l R_{ab}) q^{ab}\= 0$. Via
field equations, this condition is equivalent to $\L_\l (T_{ab}
\l^a n^b) \= 0$, where $q^{ab}$ is any ``inverse'' of $q_{ab}$.
(As in footnote 4, these conditions are independent of the choice
of $q^{ab}$ and $n^a$.) In this sub-section, we will show $\IH$
generically admits an unique $[\l]$ such that the transversal
expansion $\mu$ is ``time independent''.

Let us choose any null normal $\l$ to $\IH$ and consider the
commutator $[\L_\l , \D]$. Due to  general properties of these two
operators, there exists a tensor field $C_{ab}^c \=
C_{(ab)}^c$ on $\IH$ such that
\be \label{commute} [\L_\l , \D_a ]K_b \= C_{ab}^c K_c \ee
for any covector field $K_a$ on $\IH$. Property (\ref{horD})
implies $C_{ab}^c h_c \= 0$ for any $h_c$ defined intrinsically on
$\IH$ satisfying $\l^a h_a \= 0$.  Hence, $C_{ab}^c$ has the form:
$$ C_{ab}^c \= -\N_{ab} \l^c $$
for some symmetric tensor field $\N_{ab}$. Conditions $\D_a \l^b
\= \omega_a \l^b$ and $\L_\l \omega_b \=0$ now imply
$$\ell^a N_{ab} = 0. $$
Note that under constant rescalings $\l \mapsto c \l$, $\N_{ab}$
remains unchanged, although under a rescaling by a function, it
does change. Therefore, a natural strategy to restrict the choice
of $\l$ is to impose conditions on $\N_{ab}$. In this sub-section
we will show that, on a generic NEH $\IH$ on which the Ricci
tensor satisfies our `time-independence' condition, one can indeed
select $[\l ]$ uniquely by requiring $\N_{ab}$ to be trace-free,
i.e.,
\be \label{tfn} q^{ab} \N_{ab} \= 0\, .  \ee
(Again the condition is independent of the choice of $q^{ab}$
because $\N_{ab}\l^b \= 0$). Thus, a generic NEH $\IH$ admits a
unique WIH structure $(\IH, [\l])$ for which $\N_{ab}$ is
transverse and traceless.

Condition (\ref{tfn}) has a simple interpretation in terms of
structures introduced in Section \ref{s3.2}. Choosing $K_a = n_a$
in (\ref{commute}), with $\l^a n_a \= -1$ and $\L_\l n_a \= 0$, we
find
$$\N_{ab} = \L_\l (\D_a n_b) \equiv \L_\l \S_{ab}.$$
Thus (\ref{tfn}) requires $q^{ab}\, \L_\l \S_{ab} = 0$. If we
require $dn =0$ as in Section \ref{s3.2}, the condition reduces
to $\L_\l \mu \= 0$. Thus, our result will establish that
\textit{a generic NEH admits a unique $[\l]$ such that $(\IH,
[\l])$ is a WIH on which the transversal expansion $\mu$ is time
independent.} As remarked in Section \ref{s3.2}, this condition is
independent of the choice of $n$ satisfying (\ref{n}).

Consider any NEH $\IH$ and introduce on it a fiducial null normal
$\l$. Without loss of generality, we can assume $\l$ is
non-extremal: $\kl\not\=0$. Suppose $\N_{ab} q^{ab} \not\=0$. Our
task is to find another null normal $\l^\prime = f\l$ for which
$(\IH, [\l^\prime])$ is a WIH with $N^\prime_{ab} q^{ab} \= 0$. A
simple calculation yields
\be \label{transn} f\N^\prime_{ab} \= f \N_{ab} + 2 \omega_{(a}
\D_{b)} f \ + \D_a\D_b f. \ee
Since $N_{ab}$ and $N^\prime_{ab}$ are transversal to $\l$ and
$\l^\prime = f\l$, the functional form of $f$ is severely
constrained. Indeed, contracting (\ref{transn}) with $\l^a$, one
obtains
$$ \D_a(\L_\l f + \kl f) \= 0, $$
which can be readily solved to conclude that the $f$ we are seeking
has a specific form:
\be \label{f} f = B e^{-\kl v} + \frac{\kappa_{(\l')}}{\kl}\qquad
{\rm where}\,\, \L_\l B \= 0 \ee
Thus our task is to find a $B$ such that $\N^\prime_{ab} q^{ab} \=
0$.

Let us now introduce a covector field $n$ on $\IH$ satisfying
(\ref{n}),  contract (\ref{transn}) with the inverse metric
$\tilde{q}^{ab}$ satisfying $\tilde{q}^{ab}n_{b} \= 0$, and use
the above form of $f$. The requirement
$N^\prime_{ab} q^{ab} \= 0$ will be met if and only if
\be \label{B}[\tilde{\D}^2 + 2 \tilde{\omega}^a \D_a  + \kl\,
\tilde{q}^{ab}\tilde\S_{ab} +  \tilde{q}^{ab} \N_{ab} ]\, B \, \=
\, - \left( \frac{\kappa_{(\l^\prime)}}{\kl} e^{\kl v} \right)
 \tilde{q}^{ab} \N_{ab}   \ee
We can hope to solve this elliptic equation to determine $B$.
However, there is a potential problem in this strategy: While
there is both an explicit time dependence (through $\exp \kl v$)
on the right hand side of this equation and an implicit time
dependence  (via $\tilde{\S}_{ab}$) on the left hand side, the $B$
we are seeking is required to be time-independent. Therefore, it
is not immediately clear whether (\ref{B}) admits \textit{any}
time-independent solution $B$.

Let us therefore examine the $v$ dependence in detail.
Fortunately, under our assumption that $q^{ab}R_{ab}$ be time
independent, we have an explicit formula for
$\tilde{q}^{ab}\tilde{\S}_{ab}$ (see (\ref{sol1}))q. Substituting
it in (\ref{B}) and using $N_{ab} = \L_\l {\S}_{ab} $, we obtain
\be \label{beq} {\M}\, B :=  \left[ \tilde{\D}^2 + 2
\tilde\omega^a \tilde{\D}_a + \tilde\D^a \tilde\omega_a +
\tilde\omega_a \tilde\omega^a - \frac{1}{2} \tilde{\cal R} +
\frac{1}{2} \tilde{q}^{ab}\, R_{ab}\right]\, B \, = \,
\kappa_{(\l^\prime)} \tilde{q}^{ab}\tilde{S}^0_{ab}\ee
where $\tilde{S}^0_{ ab}$ satisfies $\L_\l \tilde{S}^0_{ab} \= 0$.
Thus, (thanks to our assumption $\L_\l (q^{ab} R_{ab}) \= 0$), the
elliptic equation on $B$ is in fact $v$-independent. Hence, we can
hope to find a time independent solution $B$ as required. By
assumption, the right side is non-zero (for, if it were zero, our
fiducial $[\l]$ already satisfies our condition (\ref{tfn})).
Therefore, if zero is \textit{not} an eigenvalue of the elliptic
operator $\M$, (\ref{beq}) is guaranteed to admit a solution
which, furthermore, is unique. One can show that the dimension of
the kernel of $\M$ is a property only of the NEH $\IH$ and does
not depend on the choice of $\l$ or of the cross-section
$\tilde{\IH}$ used to construct $\M$.

We will say that \textit{a NEH is} {generic} \textit{if the
elliptic operator} $\M$ \textit{on $\hIH$ has trivial kernel} for
\textit{some} choice of $\l$ with $\kl \not\= 0$ (and $\L_l R_{ab}
q^{ab} \= 0$ holds). Then, on generic NEHs, there is exactly one
$[\l^\prime ]$ for which $(\IH, [\l^\prime])$ is a WIH with
trace-free $\N^\prime_{ab}$ (or, time-independent transversal
expansion $\mu$). This establishes our assertion. (We will obtain
an alternate and geometrically more transparent uniqueness result
in section \ref{s6}.) Furthermore, the generic property guarantees
that $\kappa_{(\l^\prime)}$ is non-zero, i.e., the WIH so selected
is non-extremal. Finally, note that even if $\M$ has a non-trivial
kernel, the right side of (\ref{beq}) may be in the image of $\M$.
In this case, preferred $[\l ]$s would exist but would not be
unique.

\textit{Remark}: If $\IH$ is a Killing horizon for a Killing field
$[\xi]$, and we choose $[\l ] = [\xi]$, the Killing vector,  $\mu
= \tilde{q}^{ab} \N_{ab}$ is guaranteed to be time-independent.
Thus, generically, our condition (\ref{tfn}) extracts from the
Killing property just the ``right'' information to select a
canonical $[\l ]$. The detailed strategy is rather subtle. For
example, although it seems natural at first, on generic NEHs we
cannot require that all of $\tilde{\S}_{ab}$ be time-independent.
The rescaling of $\l$ provides a single free function $B$ and we
can be adjust it to make only the trace time-independent.

Note however, that the existence of a Killing vector in the
neighborhood of the horizon is not sufficient to guarantee that
$\M$ is invertible. This is in particular the case for extremal
Killing horizons $\IH_K$, for some Killing vector $\xi$ defined
near $\IH$. For, in this case, there obviously does exist a
$\l^\prime$ for which $\N^\prime_{ab}$ is trace-free (namely
$\l^\prime \= \xi$) so there is indeed a non-trivial $B$ relating
this $\l^\prime$ and the fiducial, non-extremal $\l$ we began
with. Since $\kappa_{(\l^\prime)}\= 0$, from (\ref{beq}) we
conclude that this $B$ is in the kernel of $\M$. Thus, in this
case, $\M$ is necessarily non-invertible, whence any extremal
isolated horizon is non-generic in the present terminology.

\section{Isolated Horizons.}
\label{s5}

In this section, we will strengthen the notion of isolation by
requiring the intrinsic metric $q$ and the \textit{full}
derivative operator $\D$ (rather than just the 1-form $\omega$) be
time-independent. We will first introduce the basic definition and
then isolate the free data and comment on the issue of existence
and uniqueness of an IH structure on a given NEH. These issues are
further discussed in some detail in Appendix \ref{sa1}.

\textbf{Definition:} An isolated horizon (IH) is a pair $(\IH,
[\ell])$, where $\IH$ is a NEH equipped with an equivalence class
$[\l ]$ of null normals such that
\be \label{ih} [\L_\l , \, \D] \, \= 0 \, . \ee
If this condition holds for one $\l$ it holds for all $\l$ in $[\l
]$.

Let $\IH$ be a NEH with geometry $(q,\D)$. We will say that this
geometry \textit{admits an isolated horizon structure} if there
exists a null normal $\l$ satisfying (\ref{ih}). This IH structure
will be said to be extremal if $\kl \= 0$ and non-extremal
otherwise. Intuitively, a NEH is an IH if the entire geometry
$(q,\D)$ of the NEH is `time-independent'. From the perspective of
the intrinsic geometry, this is a stronger and perhaps more
natural notion of `isolation' than that captured in the definition
of a WIH. Indeed the basic condition (\ref{wih}) in the definition
of a WIH can be reformulated as
$$ [\L_\l, \D] \l^a \= 0, $$
i.e., as restricting the action of the left side of (\ref{ih}) to
$\l$. In the terminology introduced in Section \ref{s4.2}, an IH
is a WIH on which the field $\N_{ab}$ vanishes identically; on a
WIH only $\N_{ab} \l^b = \L_\l \S_{ab} \l^b = \L_\l \omega_a$
vanishes. An IH mimics properties of a Killing horizon to a
slightly higher degree than a WIH. However, explicit examples
\cite{pc} as well as an analysis \cite{jl} using the initial value
problem based on two null surfaces \cite{fr} shows there is an
infinite-dimensional class of other examples. In particular, while
all geometric fields are time-independent on a Killing horizon,
the field $\Psi_4$, for example, can be time-dependent on a
generic IH.

We saw in Section \ref{s3} that any NEH can be made a WIH simply
by choosing an appropriate class $[\l ]$ of null normals. The
situation with IHs, by contrast, is quite different. Not every NEH
admits a null normal satisfying (\ref{ih}); this condition is a
\textit{genuine} restriction. Indeed, we saw in Section \ref{s4.2}
that, generically, weak isolation and the condition $N_{ab}q^{ab}
\= 0$ exhaust the rescaling freedom in the choice of $[\l]$. The
resulting $[\l]$ is then the only candidate for an IH structure on
the given NEH. However, in general, the resulting $[\l]$ will not
be such that the trace-free part of $N_{ab}$ also vanishes.

Given a candidate for a black hole space-time, it is generally
easy to verify whether one's guess for the horizon is in fact a
NEH, which can be readily made a WIH simply by choosing the null
normal appropriately. However, it is considerably more difficult
to verify whether there exists a null normal which makes it an IH.
Necessary conditions for the existence of such a null normal
follow from (\ref{ls1}) and some of these are discussed in
Appendix \ref{sa1}.

Let us now extract the freely specifiable `data'  on an IH. The
structure of a WIH is specified by an equivalence class $[\l]$ and
a pair $(q, \D)$, or, as in Section \ref{s3.2}, by a constant
$\kl$ and a triplet of fields $(q_{ab}, \omega_a,
\tilde{\S}_{ab})$ satisfying (\ref{wih}) and (\ref{ls1}) on $IH$.
However, since all these fields are time-independent on an
isolated horizon, they are constrained further. Let us first
consider the case when $\kl \not\= 0$. Then, (\ref{sol1}) implies
$\tilde{\S}^0_{ab}$ must vanish, whence $\tilde{\S}_{ab}$ is
completely determined by $q_{ab}, \omega_a$ and the matter fields
(ie the pull-back of the four dimensional Ricci tensor to $\IH$):
\be \label{st1} \tilde\S_{ab} = \frac{1}{\kl}\left[\tilde\D_{(a}
\tilde\omega_{b)} + \tilde\omega_a \tilde\omega_b - \frac{1}{2}
\tilde{\cal R}_{ab} + \frac{1}{2} \tilde{q}_a{}^c\,
\tilde{q}_b{}^d \, R_{cd}\right] \ee
Therefore, given $R_{\pback{ab}}$, to specify the geometry of a
non-extremal IH, following the procedure of Section \ref{s3.2},
let us fix a 2-sphere cross section $\tIH$ of $\IH$ and fields
$(q_{ab}, \omega_a)$ thereon such that: i) the pull-back
$\tilde{q}_{ab}$ of $q_{ab}$ is a positive definite metric on
$\tIH$, ii) $q_{ab}\l^a = 0$; and, iii) $\omega_a\l^a =: \kl$ is a
non-zero constant. We then \textit{define} $\tilde{\S}_{ab}$ on
$\tIH$ via (\ref{st1}) and extend all these fields to $\IH$ by
demanding they be Lie-dragged by $\ell$. The resulting triplet
$(q_{ab}, \omega_a, \tilde{\S}_{ab})$ then defines an IH geometry.
Thus, relative to the WIH case considered earlier, there is no
longer the freedom to choose the transversal expansion $\mu$ and
shear $\lambda$ on $\tIH$; these fields are completely determined
by the pair $(q, \omega)$.

To conclude, let us consider extremal IHs. If $\kl =0$,
$\tilde{\S}_{ab}$ is given by (\ref{sol2}). Isolation implies it
is time-independent. Rather than determining $\tilde{\S}_{ab}$ in
terms of $(q_{ab}, \omega_a)$ as in the case $\kl \not\= 0$, this
condition now implies that $(q_{ab}, \omega_a)$ are
\textit{themselves constrained} by
\be\label{st2}\tilde\D_{(a} \tilde\omega_{b)} + \tilde\omega_a
\tilde\omega_b - \frac{1}{2} \tilde{\cal R}_{ab} + \frac{1}{2}
\tilde{q}_a{}^c\, \tilde{q}_b{}^d \, R_{cd} \, = \, 0 , \ee
while $\tilde{\S}_{ab}$ is now free (but of course, time
independent). Therefore, in this case, the free data consists of
triplets $(q_{ab}, \omega_{a}, \tilde{\S}_{ab})$ on $\tIH$, where
$(q_{ab}, \omega_a)$ are now subject, in addition to the
conditions given above (in the $\kl \not\= 0$ case), also to
(\ref{st2}), while $\tilde{\S}_{ab}$ is only required to be
symmetric and transverse to $\ell$. In the Newman-Penrose
notation, in this case $m_a$ and $\pi$ are now subject to
(\ref{st2}) while $\mu, \lambda$ are now free.

Finally, in the non-extremal case, we can again eliminate the
(gauge) freedom in the choice of the cross-section simply by
restricting ourselves to the `good cuts' of Section \ref{s3.3}. In
this case, $\tilde\omega$ is subject to $\tilde{q}^{ab}\tilde{D}_a
\tilde{\omega}_b \= 0$. Then, given $R_{\pback{ab}}$, the problem
of specification of $(q, \D)$ reduces to that of specifying
$(\hat{q}, \hat{U})$ on the 2-sphere $\hIH$ of generators of
$\IH$, where $\hat{U}$ is the rotational scalar (see
Eq(\ref{divfree})). The complete diffeomorphism invariant
information of an IH structure is encoded in the diffeomorphism
class of fields $(\hat{q}, \hat{U})$ on $\hat\IH$. If two distinct
IH structures $(\IH, [\l], q, \D)$ and $(\IH, [\l'], q', \D')$
yield the same pair $(\hat{q}, \hat{U})$ on $\hIH$, then there is
a diffeomorphism from $\IH$ on to itself which maps $[\l]$ to
$[\l']$. In this sense, the IH structure is unique in the
non-extremal case. The only remaining question is: can a NEH admit
two distinct IH structures, one non-extremal and the other
extremal? This is possible but, as we show in Appendix \ref{sa1},
the corresponding horizon geometry is very severely constrained.

\section{Analytic extension of $(\IH, [\l], q, \D )$}
\label{s6}

In the last three sections, we presented constructions to select
canonical null normals $[\l]$ on NEHs. In this section, we will
present an alternate characterization of these canonical $[\l ]$
using an analytic extension of  $\IH$ and fields thereon. This
characterization is again geometrical and, furthermore,
considerably easier to visualize.

Let us begin with some preliminaries. Consider a NEH $\IH$ and
introduce on it an affinely parametrized null normal $\l_0 $ and
denote the compatible coordinate by $v_0$. Then $(\IH, [\l_0])$ is
a an extremal WIH. Therefore, $\L_{\l_0} q_{ab} \= 0$ \textit{and}
$\L_{\l_0} \omega_a \= 0$. Furthermore, $\tilde{\S}_{ab}$ which
captures the remaining information in $\D$ is explicitly given by
(\ref{sol2}). Since, by inspection, $\tilde{\S}_{ab}$ is analytic
in $v_0$, and $q_{ab}$ and $\omega_a$ are independent of $v_0$, we
can analytically extend $\IH$ to $\bIH$ on which $v_0$ runs from
$-\infty$ to $\infty$, and extend $(\l_0, q_{ab}, \D)$ to
$\bar\IH$. Note that we do \textit{not} assume that the space-time
is analytic even near $\IH$; we have extended $\IH$ as an abstract
3-manifold endowed with certain fields.

In this section, we will work with this analytic extension and
fields defined intrinsically thereon. The vector field $\l_0$ is
complete on $\bIH$ and so are \textit{all} other affinely
parametrized null normals. Thus, our analytic extension does not
depend on the initial choice of the extremal $\l_0$. While these
affinely parametrized null normals are nowhere vanishing,
non-extremal null normals $\l$ can not be everywhere future
pointing (see below). Therefore, in this section we will drop the
requirement that the null normals be nowhere vanishing and future
directed.

We can endow $\IH$ with \textit{any} non-extremal WIH structure
$(\IH, [\l])$ with surface gravity $\kl$ by a rescaling
\be\label{l}\l_0 \mapsto \l \= \kl (v_0 - B)\, \l_0 \ee
for some $B$ satisfying $\L_{\l_0} B \= 0$. Since the surface
gravity rescales as $\kl \mapsto c\kl $ under constant rescalings
$\l \mapsto c\l$ with $c >0$, the function $B$ is unaffected by
these rescalings; thus, there is a 1-1 correspondence between the
functions $B$ and non-extremal $[\l]$. For any given $B$, it is
clear by inspection that $[\l]$ vanishes on precisely one 2-sphere
cross-section $\tIH_\l$ of $\bIH$, given by $v_0=B$. $\tIH_\l$
will be referred to as the \textit{cross-over 2-sphere}. If $[\l]$
is future directed in the future of $\tIH_\l$, it is past directed
in the past. The 1-1 correspondence between functions $B$ and
non-extremal $[\l]$ also implies that, given \textit{any}
cross-section $\tIH$, there is a null normal $\l$ (defining a
non-extremal WIH) which vanishes on it. Thus, \textit{there is a
1-1 correspondence between equivalence classes $[\l ]$ of
non-extremal WIH null normals and 2-sphere cross-sections
$\tIH_\l$ of $\bIH$.} We will say that $[\l]$ and the cross-over
2-sphere $\tIH_\l$ it defines are \textit{compatible} with one
another.

In Section \ref{s4.2}, we introduced a strategy to select a
canonical, non-extremal $[\l ]$ on a generic NEH. It is therefore
natural to ask if a structure on the associated cross-over
2-sphere can be used to characterize this $[\l ]$. We will now
show that the answer is in the affirmative: the expansion of every
null vector field $n$ orthogonal to $\tIH_\ell$ \textit{vanishes}
on $\tIH_{\ell}$ identically. (The vanishing  is independent of
the choice of the null normal $n$ so long as it is non-zero and
finite on $\tIH_{\l}$.)

To show this, let us suppose that $(\IH, [\l])$ is a non-extremal
WIH such that $q^{ab}N_{ab} \= 0$, i.e.,  such that for every
$n_a= -\D_a v$, with ${\cal L}_\l v \= 1$,
$$ 2\mu\ :\=\ \tq^{ab}\, \D_b n_a  $$
is time-independent in the region $\IH$ of $\bIH$ on which $\l$ is
nowhere vanishing and future directed. Now, since $[\l]$ vanishes
on $\tIH_\l$, any $n$ satisfying $\l^a n_a \= -1$ must diverge on
$\tIH_\l$. Therefore, to evaluate the transversal expansion of
$\tIH_\l$, let us pass to an appropriately rescaled $n$, say
$n^0$, which does not diverge or vanish on $\tIH_\ell$. Without
loss of the generality, we may assume that $\ell$ is given by
(\ref{l}) with $B=0$, so we can choose
$$ (n^{0})_a\ \=\ \kl v_0 n_a, \quad{\rm or}\quad
n_a\ \=\ -\frac{1}{\kl v_0}\, {\D_a v_0}. $$
Then,
$$ 2\mu_0 :\= \tilde{q}_0^{ab}\, \D_a (n^0)_b \= \kl v_0 \mu  $$
where we have used the fact that $\tilde{q}_0^{ab} \=
\tilde{q}^{ab}$ (because $n$ and $n^0$ are orthogonal to the same
family of 2-spheres).  Since $(n^0)_a$ is smooth on $\bIH$, so is
$\mu_0$. But by construction $\mu$ is time independent to the
future of $\tIH_\l$ and $v_0 \=0$ on $\tIH_\l$. Therefore, we
conclude,
 $$\mu_{0}\mid_{{\tIH}_\l} \= 0.$$
Thus, \textit{the preferred cross-section singled out by $[\l]$ is
distinguished by the fact that \textit{both} of its null
expansions vanish.} This is an intrinsic property of the
cross-section; if it holds for one pair of (well-defined) null
normals, it holds for all.
\footnote{Note incidentally that, since $n$ diverges at $\tIH_\l$,
we can not conclude that $\mu$ is zero on $\tIH_\l$: In the $(\l ,
n)$ frame, it is meaningful to calculate spin-coefficients only
away from $\tIH$ and $\mu$ is time-independent only in that
region.}

Next, let us consider the converse. Let $\tIH$ be a cross-section
of ${\bIH}$ such that the expansion of every null vector field
orthogonal to $\tIH$ vanishes everywhere on $\tIH$. We will now
show that $[\ell]$ which is compatible with $\tIH$  (i.e., for
which $(\IH, [\l])$ is a non-extremal WIH and  $\ell\mid_{\tIH}
=0$) satisfies $N_{ab}q^{ab} \=0$. As before, without loss of
generality, we may assume, that
$$ v_0\mid_{\tIH}\ \=\ 0, \quad{\rm and}\quad \ell\ \=\ v_0\ell_0. $$
From (\ref{sol2}) we know that $(S^0)_{ab} :\= \D_a n^0_b =
-\D_a\D_b v_0$ satisfies
$$ {\tS}^0_{ab} \=\ (\tS^0)^0_{ab} + (\tS^0)^1_{ab}v_0 $$
where $(\tS^0)^0_{ab}$ and $(\tS^0)^1_{ab}$ are Lie dragged by
both $\ell_0$ and $\ell$. From the vanishing of the expansion of
$(n^0)_a$ on $\tIH$, we conclude
$$ \tq^{ab} (\tS^0)_{ab}\mid_{v_0=0}\ \=\ \tq^{ab} (\tS^0)^0_{ab}\
\=\ 0. $$
Now, since $v \= \frac{1}{\kl}\ln v_0$ is compatible with $\l$,
$\tS_{ab} \= -\tilde{q}_a{}^c \tilde{q}_b{}^d \D_a\D_b v$ is given
just by the rescaling,
$$ \tS_{ab}\ \=\ \frac{1}{\kl v_0}(\tS^0)_{ab}\ \=\ \frac{1}{\kl
v_0}(\tS^0)^0_{ab} + \frac{1}{\kl}(\tS^0)^1_{ab}. $$
Therefore
\be 2\mu\ \=\ \tq^{ab}\tS_{ab}\ \=\ \tq^{ab}(\tS^0)^1_{ab}, \ee
is constant along the null generators of $\IH$.

To summarize, \textit{the cross-over 2-sphere of a $[\ell]$
defining a non-extremal WIH is non-expanding in the both
orthogonal null directions if and only if $[\l]$ satisfies
$N_{ab}q^{ab}\ \=0$}. Note that, in this analysis, we did not have
to impose the `genericity' condition of Section \ref{s4.2}: it
sufficed to assume that we are given a non-extremal WIH $(\IH, [\l
])$ on which $\mu$ is time-independent.

Finally, let us consider a non-extremal IH $(\IH, [\l])$ as in
Section \ref{s5}. Repeating the above arguments but using all
components of $\tS_{ab}$ (rather than just the trace) we can
conclude that the $\tilde\S_{ab}$ vanishes on the cross-over
2-sphere $\tIH_\l$. Conversely, let us suppose that we are given a
non-extremal WIH  with a cross-section $\tIH$ on which
$\tilde\S_{ab}$ vanishes. Then the $[\ell]$ defined by $\tIH$
endows $\IH$ with the structure of a non-extremal IH. Thus, on a
WIH \textit{the cross-over 2-sphere of  $[\ell]$ is non-expanding
and shear free in the both orthogonal null directions if and only
if  $(\IH, [\ell])$ is an IH}.

\section{Discussion}
\label{s7}

In this paper, we analyzed geometrical structures defined
intrinsically on non-expanding, weakly isolated and isolated
horizons. The intrinsic geometry of a NEH is characterized by the
pair $(q, \D)$ consisting of a `metric' $q_{ab}$ of signature
(0,+,+)  and a compatible derivative operator $\D$. Given
\textit{any} null normal $\l$, this pair satisfies two equations:
 $\L_\l q_{ab} \= 0$ and $\D_a\l^b \= \omega_a \l^b$ for some 1-form
$\omega$ on $\IH$ (see Section \ref{s2}). $\omega$ is a potential
for the imaginary part of $\Psi_2$ and determines the angular
momentum of $\IH$ \cite{abl2}. A WIH is a NEH equipped with an
equivalence class $[\l]$ of null normals such that $\L_\l \omega
\= 0$, or, equivalently, $[\L_\l,\, \D] \l^a \= 0$. (Here two null
normals $\l$ and $\l^\prime$ are regarded as equivalent if they
related by a \textit{constant} rescaling.) The notion of
`isolation' provided by this condition suffices to show that the
zeroth and the first laws of black hole mechanics can be extended
to WIHs \cite{afk,abl2}. On an IH, the normals $[\l]$ are required
to satisfy a stronger condition: $[\L_\l\, , \D]\= 0$.

Thus, as we move from a NEH to an IH, additional intrinsic
geometrical structures are assumed to be time-independent. On a
NEH the intrinsic metric $q_{ab}$ is time-independent; on a WIH
the analog of the `extrinsic curvature' is also time-independent;
while on an IH the entire intrinsic geometry is time-independent.
In this sense, the three notions mimic Killing horizons to
increasing degrees, thereby capturing the notion that the horizon
is `isolated' in an increasingly stronger sense. Note however
that, while every Killing horizon is an IH, the converse is not
true. A sub-family of Robinson-Trautman space-times provides
explicit examples of space-times which admit isolated horizons but
do not admit a Killing vector in any neighborhood of it \cite{pc}.
More generally, existence theorems \cite{fr} based on two
intersecting null surfaces have been used to show that Einstein's
equations admit an infinite dimensional family of solutions with
isolated horizons which are not Killing horizons \cite{jl}.

We were able to isolate the `freely specifiable' parts of the
intrinsic geometry of WIHs and IHs and show how the remainder is
determined by Einstein's equations (possibly with matter sources).
We also compared the situation with that at null infinity and with
the standard initial value formulation on space-like surfaces (see
Sections \ref{s3.2} and \ref{s5}). These results clarify the
interplay between geometric structures and field equations. A
second and perhaps more important set of results concerns the
issue of uniqueness of null normals $[\l]$ which make a given NEH
a WIH or an IH. Given a NEH, one can \textit{always} choose a
family of null normals $[\l]$ on it to make it a WIH and
furthermore, there is an \textit{infinite}-dimensional freedom in
the choice (Section \ref{s3.1}). In this sense, the same NEH
geometry admits infinitely many WIH structures (Section
\ref{s3.1}) However, generically, one can select an equivalence
class $[\l]$ uniquely by considering any family of 2-sphere
cross-sections of $\IH$ preserved by the diffeomorphisms generated
by $\l^a$ and requiring that its transversal expansion should be
time-independent. (See Section \ref{s4.2}. In the Newman-Penrose
language, the requirement is that the spin-coefficient $\mu$ be
time-independent.) With IHs, the situation with existence is quite
different: Not every NEH geometry can admit an IH structure. Thus,
requiring that a NEH be isolated is a \textit{genuine}
restriction. Assuming that the NEH geometry does admit an IH
structure, we can ask if the corresponding $[\l]$ is unique. We
showed that for uniqueness to fail the horizon geometry has to be
\textit{very} special; in the generic case, the horizon must admit
a foliation on which shear and expansion of \textit{both} null
normals vanishes and, furthermore, the space-time curvature is
severely restricted (see Appendices). In the Kerr family
---and hence also in `nearby solutions'---  the uniqueness result
does hold. Finally, we also showed that every non-extremal WIH
admits an intrinsically defined foliation (Section \ref{s3.3}).

These results have significant applications to numerical
relativity, particularly to the problem of extracting physics from
the numerically evolved strong-field, near-horizon geometry.
Consider, for example, dynamical processes in which black holes
form or grow due to inflow of radiation and/or matter, or
coalescence. The end point of these processes is a single black
hole with matter or gravitational radiation in the exterior.
Numerical simulations indicate that, at late times,
back-scattering becomes negligible and the world tube of apparent
horizons becomes a NEH within numerical errors.  Then, using the
expressions provided by Hamiltonian techniques, one can compute
the angular momentum and the mass of these NEHs \cite{afk,abl2},
directly in terms of the physical fields \textit{defined at the
horizon}. This procedure has already been implemented in numerical
codes \cite{num}. Note that this procedure can be carried out
without having to embed the given numerical space-time in a
probable Kerr geometry, a task which is generally difficult
because one has no a priori knowledge of the Kerr metric in the
coordinates used in numerical simulations.

Next, using our results from Section \ref{s4.2}, we can
generically select an unique equivalence class $[\l ]$ of null
normals. Furthermore, using a procedure suggested again by
Hamiltonian techniques \cite{afk,abl2}, one can even eliminate the
freedom to rescale the null normal by a constant and fix the
normalization of $\l$ entirely in terms of the area and angular
momentum of the horizon. In the non-extremal case, generically
encountered, one can introduce the geometrical foliation of $\IH$
(of Section \ref{s3.3}) and using transversal geodesics
originating from points on these 2-spheres, obtain a foliation of
(the near horizon portion of) space-time by a 1-parameter family
of null hypersurfaces. These in turn enable one to introduce
preferred a null tetrad $(\l, n, m, \bar{m})$ and coordinates
$(v,r,\theta,\phi)$ in the \textit{strong field geometry} near the
horizon. Note that the construction is quite rigid: the only
freedom is to perform a $v, r$ \textit{independent} $U(1)$
rotation $m \mapsto \exp {if} m$ in the tetrad and change
coordinates through $v \mapsto v + {\rm const}$ and $(\theta,
\phi) \mapsto (\theta^\prime, \phi^\prime)$ where the primed
coordinates are independent of $v,r$. Even this remaining freedom
can be eliminated by additional geometrical prescriptions in
generic cases. Any geometrical field ---such as $\Psi_4$ in this
tetrad--- which is insensitive to this freedom is a physical
observable. Therefore, it is physically meaningful to directly
compare such quantities in \textit{distinct} numerical
simulations. In particular, these structures provide a means to
meaningfully plot wave forms even in the strong field, near
horizon geometry. Next, the past null hypersurface originating in
a 2-sphere cross-section in the distant future is likely to be an
excellent approximation to future null infinity ${\mathcal I}^+$.
Effort is under way to provide expressions of flux of energy and
angular momentum carried away by gravitational and other radiation
across such null surfaces and analyze their properties. Finally,
this framework is also being used to probe the physics of initial
data sets. For, if in the binary black hole problem the holes are
sufficiently far, one expects from post-Newtonian considerations
that the world tubes of the two apparent horizons would be well
modelled by WIHs. The Hamiltonian considerations are then again
applicable and, given the full initial data, one can calculate the
mass, linear momentum and spin of the two WIHs. Consider for
example the Brill-Lindquist \cite{bl} initial data for two widely
separated black holes so that the distance $d$ between them is
much larger than $GM_{\rm ADM}$, where $M_{\rm ADM}$ is the total
ADM mass. One finds that the individual horizon masses $M_{\IH_1}$
and $M_{\IH_2}$ are related to $M_{\rm ADM}$ via the physically
expected relation:
$$ M_{\rm ADM} = M_{\IH_1} + M_{\IH_2} - G \frac{M_{\IH_1}M_{\IH_2}}{d}
+ {\cal O}(\frac{GM_{\rm ADM}}{d})^2 . $$
Extension of this relation to more general initial data sets is in
progress . This exploration should, in particular, shed some light
on the question of `how much radiation there is' in certain
initial data sets.

In the next paper \cite{abl3}, we will use our current results to
analyze in detail the \textit{4-dimensional} geometrical
structures in space-time regions near weakly isolated and isolated
horizons. This analysis paves the way to study perturbations of
isolated horizons. A complete characterization of the Kerr
isolated horizon \cite{lp2} already exists. Therefore, study of
perturbations will also provide tools to systematically
investigate an important issue that has remained largely
unexplored: what is the precise sense in which the near horizon
geometry approaches that of Kerr space-times in physically
interesting dynamical processes?

\bigskip\bigskip

\textbf{Acknowledgments} We would like to thank Robert Beig, Badri
Krishnan, Pablo Laguna, Richard Matzner, Tomasz Pawlowski and
Deirdre Shoemaker for correspondence and discussions. This work
was supported in part by the NSF grants  PHY-0090091,
PHY-97-34871, the NSF cooperative agreement PHY-01-14375, the
Polish CSR grant 2 P03B 060 17, the guest-program of the Albert
Einstein Institute and the Eberly research funds of Penn State.

\appendix
\section{Uniqueness and existence of IH structures}
\label{sa1}

\subsection{Uniqueness}
\label{sa1.1}

At the end of Section \ref{s5}, we saw that if an NEH geometry
$(q, \D)$ admits two \textit{non-extremal} IH structures $[\l]$
and $[\l^\prime]$, then there is a geometry-preserving
diffeomorphism on $\IH$ which maps $[\l]$ to $[\l^\prime]$. For
completeness, we will now address the question: can a NEH geometry
support a non-extremal and an extremal IH structure? We will find
that this can happen only if that NEH geometry is \textit{very}
special.

Recall from Section \ref{s4.2} that if the horizon geometry is
generic, i.e., if the elliptic operator ${\bf M}$ of equation
(\ref{beq}) has trivial kernel, the condition $N_{ab}q^{ab} \= 0$
already implies that $[\l]$ is unique. In the present case, the
burden on the normal is greater; $N_{ab}$ itself has to vanish.
Therefore, we will be able to obtain a stronger uniqueness result.

Let $(\IH, [\l])$ be an IH. Then $\bar\l \= \bar{f}\l$ also
satisfies (\ref{ih}) if and only if $\bar{N}_{ab} \= 0$ which, by
(\ref{f}), is possible if and only if $\bar{f}$ satisfies:
\be \label{ihf1} \D_a \D_b \bar{f} + 2\omega_{(a} \D_{b)}\bar{f}
\=0\, . \ee
As in Section \ref{s4.2}, by transvecting this equation with
$\l^a$ we conclude that $\bar{f}$ must be of the form (see
(\ref{f}))
$$\bar{f} = B e^{-\kl v} + \frac{\kappa_{\bar\l}}{\kl}\, ,$$
where $\L_\l B \= 0$ and $\kappa_{\bar\l}$ is the surface gravity
of $\bar\l$. Substituting this form back in (\ref{ihf1}) we find
that on each 2-sphere $\tIH$ defined by $v={\rm const}$, $B$ must
satisfy
\be \label{beq2} \tilde{\D}_a \tilde{\D}_b B +2
\tilde{\omega}_{(a} \tilde{\D}_{b)}B + \kl \tilde{S}_{ab} B \= 0
\ee
where $\tilde{S}_{ab}$ is given in terms of $(\tilde{q},
\tilde\omega)$ by (\ref{st1}). Since the single function $B$ is
subject to three different equations on each $\tIH$, a non-trivial
solution will exist if and only if the coefficients $(\tilde{q},
\tilde\omega)$ are severely constrained. Indeed, our discussion in
Section \ref{s4.2} shows, generically, there is no non-trivial
solution even to the trace of this equation. In the remainder of
this sub-section we will explore these constraints under two sets
of mild assumptions.

By taking derivatives of this equation one can obtain an
integrability condition of the following form
\be \tilde{r}_a{}^b \tD_b B\ \=\ \tilde{s}_a B\label{int} \ee
with
\ba \tilde{r}_a{}^b\ :&\ \=&\ 3\te^{cd}(\tD_c\tomega_d) \tq_a{}^b
+ \frac{3}{2}\tilde{{\cal R}} \te_a{}^b -
R_{cd}\te^{cb}\tilde{q}_a{}^d
\label{r}\\
\tilde{s}_a\ :&\=&\ 2\kl \te^{dc}(\tD_c+\tomega_c)\tS_{da}\ \ \=\
\kl\tepsilon^{cd}n_b \tilde{q}_a{}^{m}R^b{}_{mcd}\label{s}\, , \ea
where
$\tepsilon^{ab}={}^{2}\!\epsilon_{cd}\tilde{q}^{ac}\tilde{q}^{bd}$
and $R^b{}_{mcd}$ is the space-time curvature. Let us now assume
that $\tilde{r}_a{}^b$ is invertible. If there are no matter
fields \textit{on} $\IH$ this condition is equivalent to assuming
that $\Psi_2$ does not vanish anywhere on $\IH$, a condition
satisfied e.g. by the Kerr family. With this assumption, we can
eliminate the derivatives of $B$ in Eq (\ref{beq2}) and obtain
\ba
&B& \,\Big[\tD_{[a}(\r_{b]}{}^cs_c) \Big]\  \=\ 0, \quad {\rm and} \quad \\
&B&\, \Big[\r_a{}^c\r_b{}^ds_cs_d + \tD_{(a}(\r_{b)}{}^cs_c) +
2\tomega_{(a}\r_{b)}{}^cs_{c} + \kl \tS_{ab} \Big]\  \=\ 0. \ea

Now, the function $B$ can not be everywhere zero on $\IH$; if it
were, we would have $\bar\l \= {\rm const}\, \l$, violating our
assumption $[\l ] \not\= [\bar{\l}]$. Therefore there is an open
region on $\IH$ on which the terms in square brackets must vanish,
thereby constraining the horizon geometry. These are severe
constraints and unless they are met the IH horizon structure is
unique. In particular, one can show that these conditions can not
be met in the Kerr family (and hence in a neighborhood of it in
the space of solutions admitting IHs).

We will now explore the constraints imposed by (\ref{beq2}) under
a different assumption. Let us suppose that $B$ is nowhere
vanishing on $\IH$ and examine restrictions on the horizon
geometry which follow from (\ref{ihf1}) and (\ref{beq2}). This is
a mild assumption. For example, if the vacuum equations hold on
$\IH$, then the integrability condition (\ref{int}) can be solved
explicitly and one can show that the solution $B$ can not vanish
at any point (see Appendix \ref{sa2}). The same conclusion holds
also in the case when the only matter field on $\IH$ is a Maxwell
field $F_{ab}$ (such that ${\cal L}_\l F \= 0$) \cite{lp}.

Let us first note that if $B$ satisfies (\ref{beq2}), then
$\bar{f} \= B \exp (-\kl v) + (\kappa_{\bar\l}/{\kl})\,$ as well
as $f^\prime \= B \exp ({-\kl v})$ satisfy (\ref{ihf1}). Since by
assumption $B$ is nowhere zero, $\l^\prime \= f^\prime \l$ is a
permissible null normal. So, using $\bar{\l} \= \bar{f} \l$ and
$\l^\prime = f^\prime \l$, we respectively obtain a non-extremal
IH $(\  IH, [\bar\l])$ \textit{and} an extremal IH $(\IH,
[\l^\prime])$ on the same horizon geometry. We will now focus on
the extremal case and exhibit the stringent restrictions on the
horizon geometry imposed by the assumption that the IH structure
is not unique.

It is straightforward to verify that $v^\prime :\= (1/\kl B)
e^{\kl v}$ is an adapted coordinate for $\l^\prime$. Set $n^\prime
\= - dv^\prime$. This $n^\prime$ satisfies our equations (\ref{n})
for $\l^\prime$ and we will denote by $\tIH^\prime$, the 2-spheres
$v^\prime = {\rm const}$ orthogonal to it. Now,
$$ \frac{1}{v^\prime}\, \, n^\prime_b \= \kl n_b + \D_b \ln B $$
and taking derivatives of both sides we obtain
$$ \frac{1}{v^\prime}\, S^\prime_{ab} + \frac{1}{{v^\prime}^2}\,
n^\prime_{a} n^\prime_{b} \= \kl S_{ab} + \D_a \D_b B. $$
Let us now use the fact that $N_{ab} \= \L_\l S_{ab} \= 0$ and
$N^\prime_{ab} =\L_{\l^\prime} S^\prime_{ab} \= 0$. This implies
$$S^\prime_{ab} \= -2 \omega^\prime_{(a} n^\prime_{b)}. $$
Transvecting this equation with $\l^\prime$, we only obtain the
identity $S^\prime_{ab} {\l^\prime}^b \= \omega^\prime_a$. But the
pull-back of the equation to the 2-spheres $\tIH^\prime$ yields
$$ \tilde{S}^\prime_{ab} \= 0\, . $$
In the Newman-Penrose notation of Appendix \ref{sa2}, we conclude
that $\rho^\prime, \sigma^\prime, \kappa_{(\l^\prime)},
\mu^\prime, \lambda^\prime$ all vanish and $m_a, \pi^\prime$ are
constrained
by (\ref{st2}). %
\footnote{If the only matter field on $\IH$ is a Maxwell field,
one can show that these conditions imply that $(q,\omega^\prime)$
are necessarily those of the extremal Kerr-Newman space-time
\cite{lp}.} %
This is a very strong restriction on the horizon geometry. For, it
implies that: i) the horizon can be foliated by a family of
2-spheres $\tIH$ for which expansions and shears of \textit{both}
families of orthogonal null normals vanish; and, ii) $(q,
\tilde\omega)$ are severely constrained by (\ref{st2}).

To summarize, a NEH geometry admitting two inequivalent IH
structures is severely restricted. Under two different sets of
mild assumptions we exhibited these restrictions explicitly. The
second set has a simple geometrical interpretation.

\subsection{Existence of an IH structure.}
\label{sa1.2}

As noted in Section \ref{s5}, not all NEHs admit an IH structure,
i.e., a null normal $\l$ such that $[{\cal L}_\l , \D] \= 0$. In
this sub-section, we will exhibit some conditions that the
geometry $(q, \D)$ of an NEH must satisfy for such an $\l$ to
exist. Although the geometrical meaning of these conditions is not
transparent, they serve to bring out the non-triviality of the
passage from WIHs to IHs. They also provide a practical tool to
show that a given NEH does not admit any IH structure.

Let $\IH$ be an NEH. To determine if it admits an IH structure, as
in Section \ref{s4.2}, one can just construct the $[\l ]$ which
endows $\IH$ with a preferred WIH structure and ask if it
satisfies (\ref{ih}). However, this criterion is often not useful
in practice because, given a specific WIH, it may not be possible
to find explicitly the canonical $[\l ]$ required in this
construction. In this sub-section, we will derive a set of
conditions which must be satisfied by \textit{any} non-extremal
WIH structure $(\IH, [\l ], q, \D)$ if there is to exist
\textit{some} null normal $\l^\prime $ such that $(\IH,
[\l^\prime], \D)$ satisfies (\ref{ih}). (See Appendix \ref{sa2}
for further details in the N-P notation).

Let $(\IH,[\ell])$ be a WIH of $\kl\not\=0$ and suppose the
geometry of $\IH$ admits an non-extremal IH $[\ell']$. As before,
\be \ell'\, \= \, (Be^{-\kl v}+\frac{\kappa_{(\ell')}}{\kl})\ell,
\ee
where ${\cal L}_\l B \= 0$ and $v$ is compatible with $\ell$. It
follows from (\ref{transn}) and $[\L_{\ell'},\D_a] \=0$, that $B$
satisfies the following system of equations
\be \label{exist} \Big[\tilde{\D}_a \tilde{\D}_b  +2
\tilde{\omega}_{(a} \tilde{\D}_{b)} + \kl \tilde{S}^{1}_{ab}\Big]
B - \kappa_{(\ell')}\tilde{S}^0_{ab} \= 0, \ee
where $\tilde{S}_{ab} \=\tilde{S}^0_{ab}e^{\kl v}+
\tilde{S}^{1}_{ab}$, and $\tilde{S}^0_{ab}$ and $\tilde{S}^1_{ab}$
are Lie dragged by $\l$ and given by (\ref{sol1}). Again, since
these are three differential equations on a single function $B$, a
non-trivial solution can exist only if the coefficients are
suitably constrained.

An integrability condition can be derived by acting on
(\ref{exist}) by $\tD_c$:
\be \tilde{r}_a{}^b \tD_b B\ \=\ \tilde{s}^1_a B -
\tilde{s}^0_a\label{int2} \ee
where $\tilde{r}_a{}^b$ and $\tilde{s}_a$ are defined in (\ref{r})
and (\ref{s}), and  $\tilde{s}^0_a e^{\kl v}\ +  \tilde{s}^1_a\
:\= \tilde{s}_a$. Substituting for $\tD_a B$ in (\ref{exist}) we
obtain two necessary conditions in which $B$ appears only
algebraically,
\ba B \Big[\tD_{[a}\r_{b]}{}^c \tilde{s}^1_c \Big] -
\tD_{[a}\r_{b]}{}^c
\tilde{s}^0_c) + \r_a{}^c\r_b{}^d s^1_{[c}s^0_{d]}\ &\=&\ 0,\label{det1}\\
B \Big[\r_a{}^c\r_b{}^d \tilde{s}_c \tilde{s}_d +
\tD_{(a}(\r_{b)}{}^c \tilde{s}^1_c) +
2\tomega_{(a}\r_{b)}{}^c \tilde{s}^1_{c} + \kl\tS^1_{ab} \Big] &-&\nonumber\\
\tD_{(a}(\r_{b)}{}^c \tilde{s}^0_c) -
2\tomega_{(a}\r_{b)}{}^cs^0_{c} + \r_{(a}{}^c\r_{b)}{}^d
\tilde{s}^1_c \tilde{s}^0_d - \klp S^0_{ab} \ &\=&\ 0.\label{det2}
\ea
Thus, if an IH structure is to exist, these equations must admit a
solution for a nowhere vanishing function $B$. Since each equation
is of the type $B\alpha + \beta \= 0$ where the coefficients
$\alpha$ and $\beta$ depend only on the geometry $(q,\D)$, the
geometry $(q,\D)$ of the IH is constrained.

\section{Main results in the Newman-Penrose notation}
\label{sa2}

Since the Newman-Penrose framework \cite{np1} is geared to null
surfaces, it is well-suited for detailed calculations involving
the three types of horizons considered in this paper. We chose not
to use it in the main body of the paper only because the
structures of interest refer only to a null vector $\l^a$ rather
than to a full null tetrad, whence expansions of geometric fields
in null tetrads can obscure the underlying covariance. However,
for the convenience of readers more familiar with the
Newman-Penrose framework, in this appendix we will translate our
main results to that notation.

To prevent a proliferation of symbols, we will use the same symbol
to denote co-vectors and their pullbacks from the space-time onto
$\IH$; the context will make it clear which of the two
possibilities is intended.

\subsection{Null surface geometry}

Let us begin with a general null surface $\IH$, not necessarily a
non-expanding horizon (NEH). A quadruple of vectors
$(m^a,\bar{m}^a,n^a,\ell^a)$ will be said to be a \textit{null
tetrad} if the only non-vanishing scalar products of its elements
are
$$ m^a \bar{m}_a\ =\ 1\ =\ -\ell^a n_a, $$
where, $m$ is complex valued and $n$, $\ell$ are real. Following
the Newman-Penrose notation, we will denote the directional
derivatives along null tetrads by
\be\label{operators} \delta=m^a\partial_a, \quad
D=\ell^a\partial_a,  \quad \Delta=\n^a\partial_a. \ee
The dual co-frame is  given by $(\bar{m}_a, m_a, -\ell_a, -n_a)$.
We will assume that the vector field $\ell^a$ is tangent to $\IH$.
It then follows that $\Re m^a$ and $\Im m^a$ are also tangential
and the pullback of $\ell_a$ to $\IH$ vanishes. The vectors
$(m^a,\bar{m}^a,\ell^a)$ span the tangent space to $\IH$  while
the dual co-frame is given by the pullbacks of $(\bar{m}_a, m_a,
-n_a)$.

In terms of the null tetrad, the degenerate metric tensor induced
on $\IH$, is given by
\be\label{q} q_{ab}\ =\ m_a\bar{m}_b + \bar{m}_a m_b, \ee

\subsection{Non-expanding horizons}

On a non-expanding horizon, the expansion and shear of $\l$ (i.e.,
the Newman-Penrose spin coefficients $\rho$ and $\sigma$) vanish.
As a result, as explained in Section \ref{s2.1}, the space-time
derivative operator $\nabla$ compatible with the 4-metric $g$
induces an intrinsic covariant derivative operator $\D$ on $\IH$.
Being intrinsic to $\IH$, it is completely defined by its action
on $\l^a, n_a$ and $m_a$. In the Newman-Penrose framework this
action can be expressed explicitly as:
\ba
\D_a\ell^b&\=& \omega_a\ell^b \label{Dl} \\
\bar{m}^b\nabla_a n_b &\=& \lambda m_a  + \mu \bar{m}_a -\pi n_a\label{mDn}\\
m^b\nabla_a \bar{m}_b &\=& -(\alpha - \bar{\beta})m_a
+(\bar{\alpha}-\beta)\bar{m}_a +(\epsilon -\bar{\epsilon})n_a \=
-\bar{m}^b\D_am_b\label{mbdm}\, . \ea
Here, all spin coefficients are complex and the 1-form $\omega_a$
is expressed in terms of them via
\be \label{omega} \omega_a\ \=\ (\alpha+\bar{\beta})m_a +
(\bar{\alpha}+\beta)\bar{m}_a -(\epsilon +\bar{\epsilon})n_a ,\ee
The fact that $\D$ is compatible with $q_{ab}$ follows from the
fact that $m^b\D_a\bar{m}_b$ is imaginary, while the torsion-free
property, $\D_{[a} \D_{b]} f \= 0$, of $\D$ can expressed via
\ba \delta\bar{\delta} - \bar{\delta}\delta\ &\=&\
(\mu-\bar{\mu})D -(\alpha-\bar{\beta})\delta +
(\bar{\alpha}-\beta)\bar{\delta}
\label{commbm}\\
 \delta D - D\delta \ &\=&\ (\bar{\alpha}+\beta -\bar{\pi})D
- (\epsilon - \bar{\epsilon})\delta \label{comDm}
 \ea

\subsubsection{\it Constraint Equations}

The pullback $R_{\pback{ab}}$ of the space-time Ricci tensor to
$\IH$ is completely determined by $q$ and $\D$:
\ba \Phi_{00} := \frac{1}{2}R_{\ell\ell}\ &\=& 0
%D\rho-\rho^2-\sigma\bar{\sigma}-(\epsilon+\bar{\epsilon})\rho\
\label{Rll}
\\
\Phi_{10} := \frac{1}{2}R_{\ell \bar{m}} \ &\=&\ D\alpha -
\bar{\delta}\epsilon - (\bar{\epsilon}-2\epsilon)\alpha +
\epsilon\bar{\beta} -(\epsilon)\pi\ \label{Rlmb}
\\
\Phi_{20} := \frac{1}{2}R_{\bar{m}\bar{m}}\ &\=&\ D\lambda -
\bar{\delta}\pi -\pi^2 -(\alpha-\bar{\beta})\pi
+(3\epsilon-\bar{\epsilon})\lambda\ \label{Rmbmb}
\\
\Phi_{11}+ \frac{1}{8}R:= \frac{1}{2}R_{m\bar{m}}\ &\=&\ D\mu
-\delta\pi -\pi\bar{\pi} +(\epsilon +\bar{\epsilon})\mu
+(\bar{\alpha}-\beta)\pi \nonumber\\
&+& \delta\alpha-\bar{\delta}\beta -\alpha\bar{\alpha}-
\beta\bar{\beta}+2\alpha\beta -\epsilon(\mu-\bar{\mu}) \
\label{Rmmb} \ea
where $R$ is the space-time Ricci scalar.

The energy condition iii) in Definition 1 of a NEH implies
$R_{ab}\l^a\l^b \= 0$ and $R_{ab} \l^a \bar{m}^b \= 0$. The first
of these was ensured in (\ref{Rll}) by the fact that the shear and
divergence of $\l$ vanish on $\IH$. The second equation imposes
restrictions on spin-coefficients. To simplify it, let us choose
the tetrad vector $m^a$ such that
\be\label{impsilon} \epsilon \= \bar{\epsilon},\ee
This `gauge choice' can \textit{always} be made and \textit{we
will employ it from now on}. Then, using the torsion-free
conditions (\ref{commbm}) and (\ref{comDm}) satisfied by $\D$, the
second restriction, $R_{ab}\l\bar{m} \= 0$ can be expressed as:
\be\label{zeroth} D(\alpha +\bar{\beta}) - 2 \bar{\delta}\epsilon
\= 2 \epsilon (\pi -\alpha - \bar\beta). \ee

\subsection{Weakly isolated horizons}

Let us now assume that $\ell$ is so chosen that $(\IH, [\l])$ is a
weakly isolated horizon (WIH). Then, it immediately follows that
\footnote{The only departure from the standard Newman-Penrose
notation we make is to denote surface gravity $(\epsilon+
\bar\epsilon)$ by $\kl$, following the convention of black hole
mechanics. This should not cause any confusion because the fact
that $\l$ is necessarily geodesic on a null surface implies that
the Newman-Penrose $\kappa$ vanishes identically.}
\be\label{kappa} \kl := 2 \epsilon  \= \const. \ee
Next, let us choose $n$ as follows:

$$ n\ \= - \ dv,\ \ {\rm where}\ \ Dv\ \=\ 1. $$
In this gauge
\be\label{mupi} \mu\ \=\ \bar{\mu}, \ \ \pi\ \=\ \alpha +
\bar{\beta}. \ee
The projection operator onto the leaves of the foliation $v \=
{\rm const}$, can be expressed as
\be \tilde{q}^a{}_b\ \=\ m_b\bar{m}^a + \bar{m}_b m^a. \ee
and the projected fields $\tilde\omega_a$ and $\tilde{\S}_{ab}$ of
Section \ref{s3.2} can be expressed as
\ba \tilde{\omega}_a\ :\=&{}& \tilde{q}^b{}_a\omega_b\ \=\
(\alpha+\bar{\beta})m_a +
(\bar{\alpha}+\beta)\bar{m}_a\label{tildeomega}\\
\tilde{\S}_{ab}:\=&{}&\tilde{q}^c_a\tilde{q}^d_b \D_c n_d\ \=
\mu(m_a\bar{m}_b +\bar{m}_am_b) + \lambda m_am_b +
\bar{\lambda}\bar{m}_a\bar{m}_b.\label{tildeN} \ea

\subsubsection{Constraints (\ref{Rmbmb}) and (\ref{Rmmb})}

Since the pull-backs of the space-time Ricci tensor to $\IH$ is
determined completely by the geometry $(q, \D)$ of the WIH, via
field equations the stress-energy tensor at the horizon constrains
this geometry. (The other components of the Ricci tensor involve
new information, not contained in $(q, \D)$ and therefore do not
impose any such restrictions.) We have already analyzed the
consequences of (\ref{Rll}) and (\ref{Rlmb}). We will now analyze
the constraints imposed by the remaining two equations. They
determine the evolution of $\tilde{\S}_{ab}$ along the null normal
$\l^a$:
\be \Lie_\ell \tilde{\S}_{ab}\ \=
(D\mu)\,(m_a\bar{m}_b+\bar{m}_am_b)+ (D\lambda)\, m_am_b +
(D\bar{\lambda}) \,\bar{m}_a\bar{m}_b.\ee
Because $\l$ is also tangential to $\IH$, this `evolution'
equation is in fact a constraint. By decomposing this equation in
to various components, we obtain:
\ba D\mu &\=& -\kl \mu + {1\over 2} \Big(\hdiv \tilde{\omega} +
2\pi\bar{\pi}-K + 2\Phi_{11} +\frac{1}{4}R \Big),\label{Dmu1}\\
D\lambda &\=& -\kl \lambda +
\mb \pi +(\a-\bar{\b})\pi +\pi^2 + \Phi_{20},\label{Dlambda1}\\
\ea
where,
\ba K &\=& \m (\a - \bar{\b}) + \bar{\m} (\bar{\a} - {\b})  -
2  (\a - \bar{\b}) (\bar{\a} - {\b}),\label{K}\\
\hdiv \homega &\=& \m(\pi) + \mb(\bar{\pi}) - (\a
-\bar{\b})\bar{\pi} -
(\bar{\a} - \b)\pi.\label{div}\\
\ea
($K$ is the Gauss curvature of $\tilde{q}_{ab}$). Finally, in the
Newman-Penrose notation, the remaining components of the pullback
$R^a{}_{\pback{bcd}}$ of the space-time Riemann tensor onto $\IH$
are given by
\ba \Psi_0\ \=\ 0, \ \ \Psi_1\ \=\ 0\\
\Psi_2 +\frac{R}{12} \ \=\ D\mu +\kl\mu -\delta\pi
+(\bar{\alpha}-\beta)\pi-\pi\bar{\pi} \label{Psi2}\\
\Psi_3\ -\ \Phi_{21}\ \=\ \mb \mu - \m \lambda + \pi \mu + \lambda
(\bar{a}- 3\b).\label{Psi3} \ea
where
\ba \Psi_0\ =\ C_{\a\b\g\d}\ell^\a m^\b\ell^\g m^\d,\ \
\Psi_1\ =\ C_{\a\b\g\d}\ell^\a n^\b\ell^\g m^\d,\\
\Psi_2\ =\ \frac{1}{2}C_{\a\b\g\d}\ell^\a n^\b\big( \ell^\g n^\d -
m \bar{m}\big),\\
\Psi_3\ =\ C_{\a\b\g\d}n^\a \ell^\b n^\g \bar{m}^\d,\ \ \Phi_{21}\
=\ \frac{1}{2}R_{ab}n^a\bar{m}^b. \ea

\subsubsection{Good cuts and the canonical WIH structures}

Given a non-extremal WIH $(\IH,[\ell])$, a representative null
normal $\l$, as shown in Section \ref{s3.3} we can obtain a
preferred foliation of $\IH$. The leaves of this foliation are
called \textit{good cuts} of $\IH$. Let us we label these cuts by
$v = {\rm const}$ with $Dv \= 1$ and set $n = -dv$. Then, in the
Newman-Penrose notation, these cuts are characterized by the
following equations:
\be \hat{\div}\homega\ \= \ \m(\pi) + \mb(\bar{\pi}) - (\a
-\bar{\b})\bar{\pi} - (\bar{\a} - \b)\pi\ \=\ 0 \quad {\rm and}
\quad  \pi\ \=\ -i\mb U. \ee

Next, given  a NEH $\IH$, the strategy of Section \ref{s4.2} is to
select a canonical $[\ell]$ by requiring that $(\IH, [\l])$ be a
WIH satisfying
\be D\mu\ \=\ 0. \ee
Our main result of Section \ref{s4.2} can be stated as follows: if
the operator
\be \M \ \= \ \m\mb + \mb\m - (\alpha-\bar{\beta})\m -
(\bar{\alpha}-\beta)\mb + \hdiv \homega + 2\pi\m + 2\bar{\pi}\mb +
2\pi\bar{\pi} - K + 2\Phi_{11} +\frac{1}{4}R, \ee
has a trivial kernel, then the canonical $[\l]$ exists, is
non-extremal and unique. In this case, $\mu$ is determined by
other horizon fields via:
\be \mu\ \=\ {1\over 2\kl} \Big(\hdiv \tilde{\omega} +
2\pi\bar{\pi}-K + 2\Phi_{11} +\frac{1}{4}R\Big). \ee

\subsection{Implications of non-unique IH structures}

\subsubsection{The general case}

Let $(\IH, [\ell])$ be an IH. We now assume that the underlying
horizon geometry $(q,\D)$ admits a distinct IH structure $[\l']$
and investigate the consequences of this non-uniqueness. In this
case have:
\be \ell'\ =\ f\ell, \ee
and using only the WIH properties of $\ell$ and $\ell'$ we know
that,
\ba
 f\ &\=&\ Be^{-\kl v} + \frac{\klp}{\kl} \ {\rm if}\  \kl\not\=0 \\
{\rm and}\ \ f\ &\=&\ \klp v - B \ {\rm if}\ \kl = 0,\ea
where $Dv\ \=\ 1$ and $ DB\ \=\ 0$. Let us first consider the case
when $\kl \not\= 0$. Then, the condition $[\L_\ell,\D] \=0 \=
[\L_{\ell'},\D]$ is equivalent to the following set of equations
on a cross-section $\tIH$ of $\IH$:
\ba &\Big[&\frac{1}{2}(\m\mb + \mb\m - (\alpha-\bar{\beta})\m -
(\bar{\alpha}-\beta)\mb) + \pi\m + \bar{\pi}\mb + \mu \kl \Big]B\
\=\ 0 \label{B1}\\
&\Big[&(\m + \bar{\alpha} -\beta + 2\bar{\pi})\m +
\bar{\lambda}\kl \Big]B\ = \ 0.\label{B2} \ea
on a cross-section $\tIH$ of $\IH$. These are equivalent to Eq
(\ref{beq2}) in the main text

The integrability conditions of this set are given by Eq
(\ref{int}), namely $\tilde{r}_a{}^b \D_b B \= \tilde{s}_a B$. In
the Newman-Penrose notation, we have
\ba \tilde{r}_a{}^b\ &\=&\ 2i \Big[ \Phi_{20}m_am^b -
\Phi_{02}\bar{m}_a\bar{m}^b - (3\Psi_2 - 2\Phi_{11}) m_a\bar{m}^b
+ (3 \overline{\Psi_2} - 2\Phi_{11})\bar{m}_am^b\Big]\label{rNP}\\
\tilde{s}_a\ &\=&\ 2i\Big[(\Psi_3 - \Phi_{21})m_a -
(\overline{\Psi_3}-\Phi_{12})\bar{m}_a\Big]. \label{sNP} \ea
and (\ref{int}) reduces to:
\be\label{intA1} (3\Psi_2\ -\ 2\Phi_{11}) \mb B - \Phi_{20}\m B\
=\ -\kl (\Psi_3 -\Phi_{21})B. \ee
In the main text we used the inverse of $\r_a{}^b$ to express
$\D_a B$ in terms of $B$ and the horizon geometry. This expression
simplifies if we make a mild assumption on matter fields on $\IH$,
namely,
\be \Phi_{20}\ \=\ 0, \ee
which automatically holds, in particular, in the electrovac case.
Then the matrix $\tilde{r}_a{}^b$ is diagonal in the null frame,
and invertible at any point of $\IH$ at which
\be 3\Psi_2 - 2\Phi_{11}\ \not\= 0 .\ee
At these points, $\r_a{}^b\tilde{s}_b = \bar\Psi m_a + \Psi
\bar{m}_a$ where
\be\label{Psi} \bar{\Psi}\ :\=\ - \frac{\Psi_3 -\Phi_{21}}{3\Psi_2
- 2\Phi_{11}}. \ee
Thus, if $\Phi_{20} \= 0$, the restriction that $\tilde{r}_a{}^b
(x)$ be invertible reduces to the condition that $3\Psi - 2
\Phi_{11}$ be non-zero at $x$. On the part of $\IH$ on which
$\Psi$ is well-defined, the integrability conditions imply
\be \label{A54} B\, \Im\, (\m +\beta -\bar{\alpha})\bar{\Psi} \
\=\ 0.\ee
Now, $B$ can not vanish identically; if it did, $\l^\prime = {\rm
const}\, \l$, contradicting our assumption $[\l] \not\= [\l']$. On
the portion of $\IH$ where $B$ does not vanish, the horizon
geometry is constrained.

So far we have focussed on the integrability conditions for
(\ref{B1}) and (\ref{B2}). These latter equations themselves
impose further constraints on the horizon geometry. Using the
integrability conditions to substitute for derivatives of $B$ in
terms of $B$ in these equations, we obtain:
\ba B\Big[(\bm + \bar{\beta}-\alpha)\Psi\  +\pi\Psi
+\bar{\pi}\bar{\Psi}
+ \kl\Psi\bar{\Psi} +  \mu\Big]\ &\=&\ 0,\label{nonu1}\\
{\rm and}\quad B\Big[(\delta + \bar{\alpha}  - \beta)\Psi\
+2\bar{\pi}\Psi + \kl\Psi^2 + \bar{\lambda}\Big]\ &\=&\
0\label{nonu2} \ea
Since $B$ can not vanish identically, the horizon geometry which
(together with the tetrad) determines the terms in the square
brackets is constrained.

Finally, let us consider the case when $\kl \=0$. Then, (\ref{B1})
and (\ref{B2}) are replaced by
\ba \Big[\frac{1}{2}\m\mb + \mb\m - (\alpha-\bar{\beta})\m -
(\bar{\alpha}-\beta)\mb + \pi\m + \bar{\pi}\mb\Big]B\ &\=&\ - \mu
\klp \label{mu}\\
{\rm and} \quad \Big[(\m + \bar{\alpha} -\beta + 2\bar{\pi})\m
\Big]B &\=&\ - \bar{\lambda}\klp\label{lam}; \ea
and the integrability conditions (\ref{intA2}) are replaced by
\be (3\Psi_2\ -\ 2\Phi_{11}) \mb\, B - \Phi_{20}\m\, B\ \=\ -\klp
(\Psi_3 -\Phi_{21}) B.\label{intA2} \ee
Assuming $\Phi_{20} \= 0$ and $\Psi$ is well-defined, $B$ again
satisfies (\ref{A54}) if $\kappa_{(\l')} \not\=0$, but
\be B \= {\rm const} \ee
if $\kappa_{(\l')} \=0$. Finally, if $\kappa_{(\l^\prime)} \not=
0$ equations (\ref{mu}) and(\ref{lam}) imply
\ba B \Re \Big[(\delta + \beta - \bar{\alpha} +\klp\Psi+
2\bar{\pi})\bar{\Psi}\Big] + \mu\ &\=&\ 0\\
B\Big[(\delta + \bar{\alpha} - \beta + 2\bar{\pi}+\klp\Psi)
\Psi\Big] + \bar{\lambda}\ &\=&\ 0\\
B \Im \Big[\delta +\beta -\bar\alpha\Big]\, \bar\Psi &\=& 0. \ea

To summarize, if the IH horizon structure is not unique, the
horizon geometry is constrained both in the non-extremal and
extremal cases. We exhibited these constraints under the
assumption that $\Phi_{20} \= 0$ and $\Psi$ is well-defined.

\subsubsection{Simplifications in the non-extremal, vacuum case.}

Let us now suppose $R_{\pback{ab}} \= 0$ and $\kl \not\=0$ on
$\IH$. We will now show that in this case $B$ and $\Psi_2$ can not
vanish anywhere (whence $\Psi$ is well-defined everywhere) on
$\IH$. Under our present assumptions, Bianchi identities yield:
\be \kl\Psi_3\ \=\ (\mb +3\pi)\Psi_2. \ee
The integrability condition (\ref{intA1}) is equivalent to
\be\label{0} B^3 \Psi_2\, e^{-3iU}\ \= \ C \ee
where $C$ is a constant. If $C \not\= 0$, $B$ and $\Psi_2$ must be
everywhere non-vanishing as we wished to show. Let us therefore
consider the other case, $C = 0$. Now, $B$ can not vanish
identically; if it did $[\l ] = [\l']$, contradicting our initial
assumption. If $C=0$, Eq (\ref{0}) implies that $\Psi_2$ must
vanish on the region on which $B$ is non-zero and  Eq (\ref{B1})
implies
\be \hDelta B\ =\ 0, \ \ {\rm whenever}\ \ \Psi_2\ =\ 0. \ee
Let $\bar{S}$ be the closure of the support of $B$. Since $\hDelta
B =0$ on $\bar{S}$ and $B$ vanishes on the boundary of $\bar{S}$,
$B$ must vanish on $\bar{S}$. This implies $B$ vanishes everywhere
on $\hIH$, contradicting our assumption. Thus, the constant $C$
can not be zero.

To summarize if $\kl \not\=0$ and $R_{\pback{ab}} \= 0$, then the
assumptions $\Phi_{20} \=0$ of the last sub-section is trivially
satisfied and, furthermore, $\Psi$ is well-defined globally on
$\IH$. In this case, the geometry is severely constrained and, if
it exists, $B$ is given by (\ref{0}).

\subsubsection{The non-extremal, vacuum, non-rotating case.}

Let as apply the above results of the last subsection to
non-rotating horizons, i.e., horizons satisfying
\be \tomega_a\ \=\ 0. \ee
Note that the intrinsic metric of these horizons need not be
spherical; arbitrary distortions are permissible. In this case, $-
\Psi_2 = K$ whence (\ref{0}) implies
\be B\ \= \ B_0 K^{-\frac{1}{3}} \ee
where $B_0$ is a constant. Integrating equation (\ref{B1}) on
$\hIH$ and substituting for $\mu$ from (\ref{st1}) we conclude
\be (\hDelta - K)K^{-\frac{1}{3}} \= 0 . \ee
Finally, using the fact that the image of the Laplace operator is
orthogonal to the constant function, we find
\be 0\ \=\  B_0 \int_{\hIH} (\hDelta -
K)K^{-\frac{1}{3}}\hat{\epsilon}\ =\ -B_0 \int_{\hIH} \big(
K^{\frac{1}{3}}\big)^2\hat{\epsilon}\ \ee
Now, because of our assumptions $K$ is nowhere vanishing, whence
the integral is positive definite. This implies $B_0\=0$ and
therefore $[\l ] \= [\l']$. Thus, in the non-extremal,
non-rotating, vacuum case, we conclude that \textit{if a NEH
admits an IH structure, that structure is unique.}

\subsection{The existence conditions.}

Finally, we will recast the discussion of Appendix \ref{sa1.2} in
the Newman-Penrose language. Let $\IH$ be a NEH. Choose
\textit{any} null normal $\l$ which endows it the structure of a
non-extremal, weakly isolated horizon and denote by $v$ a
compatible coordinate. Suppose $\ell'$ is another null normal
which defines an IH structure on $\IH$. Then, $\ell'=Be^{-\kl v} +
\frac{\klp}{\kl}$ (with $DB=0$) and the geometry satisfies the
conditions (\ref{exist}). In the Newman-Penrose notation, they
read,
\ba \frac{1}{2}\Big[ \mb\m + \m\mb +(2\pi - \alpha +
\bar{\beta})\m + (2\bar{\pi} -\bar{\alpha} +\beta)\mb +
2\kl \mu^1\Big] B\ & \=& \ \klp\mu^0\\
{\rm and} \quad \Big[\mb\mb +(2\pi+\alpha-\bar{\beta})\mb + \kl
\lambda^1 \Big]B\ & \=& \ \klp\lambda^0, \ea
where $\mu\ \=:\ \mu^0 e^{-\kl v} + \mu^1$ and $ \lambda\ \=:\
\lambda^0 e^{-\kl v} + \lambda^1$ with $D\mu^0 \=D\mu^1
\=D\lambda^0 \=D\lambda^1 \=0$.  The $\tilde{r}_a{}^b$ and
$\tilde{s}_a$ used in the integrability condition (\ref{int2}) are
expressed in the Newman-Penrose formalism in Eqs (\ref{rNP}) and
(\ref{sNP}). Using these expressions, the integrability condition
reads
\ba (3\Psi_2\ -\ 2\Phi_{11}) \mb B - \Phi_{20}\m B\ \=\ -\kl
(\Psi_3 -\Phi_{21})^1 B + \klp(\Psi_3 -\Phi_{21})^0,\\
{\rm and} \quad (\Psi_3 -\Phi_{21})\ \=:\ (\Psi_3 -\Phi_{21})^0
e^{-\kl v} + (\Psi_3 -\Phi_{21})^1, \ea
where $D(\Psi_3 -\Phi_{21})^0\=D(\Psi_3 -\Phi_{21})^1\=0$. Let us
write down the equations  (\ref{det2}) and (\ref{det1}) assuming
again $\Phi_{02}\=0$, thereby making the matrix $\tilde{r}_a{}^b$
diagonal in the null frame. The function $\Psi$ of (\ref{Psi})
also has the form
\be \Psi\ :\=\ \Psi^0  e^{-\kl v} + \Psi^1, \ee
where $D\Psi^0 \=D\Psi^1\ \=0.$  Then, wherever
$3\Psi_2-2\Psi_{11}\not\=0$ the equations (\ref{det2}) and
(\ref{det1}) hold and they read,
 \ba \Im\Big[B(\mb
-\alpha +\bar{\beta}) \Psi^1 - \frac{\klp}{\kl}(\mb -\alpha
+\bar{\beta})\Psi^0 - \klp \Psi^1\overline{\Psi^0}\Big]\ &\=&\ 0\\
\Re\Big[B\Big((\mb +2\pi - \alpha +\bar{\beta}) \Psi^1
+ \kl |\Psi^1|^2 + \mu^1\Big)-&{}&\nonumber\\
\frac{\klp}{\kl}(\mb + 2\pi -\alpha +\bar{\beta})\Psi^0
-\kl\overline{\Psi^1}\Psi_0-\frac{\klp}{\kl}\mu^0
 \Big]\ &\=&\ 0,\\
B\Big((\mb + 2\pi +\alpha-\bar{\beta})\overline{\Psi^1}
+ \kl \overline{\Psi^1}^2 + \lambda^1 \Big)&{}&\nonumber\\
+ \klp\overline{\Psi^1}\overline{\Psi^0} -\frac{\klp}{\kl}(\mb +
2\pi +\alpha-\bar{\beta})\overline{\Psi^0}
-\frac{\klp}{\kl}\lambda^0\ &\=&\ 0. \ea

These are a set of necessary conditions on the horizon geometry
for an IH structure to exist.

\end{document}